\newcommand*{\ATLASLATEXPATH}{}
\documentclass[PAPER, texlive=2016, UKenglish, cernpreprint]{\ATLASLATEXPATH atlasdoc}
 
\usepackage[subcaption]{\ATLASLATEXPATH atlaspackage}
\usepackage{\ATLASLATEXPATH atlasbiblatex}
 
\usepackage{\ATLASLATEXPATH atlasphysics}
 
\graphicspath{{logos/}{figures/}}
 
\usepackage{ANA-SUSY-2019-08-PAPER-defs}
 
\usepackage{float}
\floatstyle{plaintop}
\restylefloat{table}

\AtlasTitle{Search for direct production of electroweakinos in final states with one lepton, missing transverse momentum and a Higgs boson decaying into two $b$-jets in \(pp\) collisions at $\sqrt{s}=13$~TeV with the ATLAS detector}
 
\AtlasJournal{EPJC}
\PreprintIdNumber{CERN-EP-2019-188}
\AtlasAbstract{
The results of a search for electroweakino pair production $pp \rightarrow \tilde\chi^\pm_1 \tilde\chi^0_2$ in which the chargino ($\tilde\chi^\pm_1$) decays into a $W$ boson and the lightest neutralino ($\tilde\chi^0_1$), while the heavier neutralino ($\tilde\chi^0_2$) decays into the Standard Model 125 GeV Higgs boson and a second $\tilde\chi^0_1$ are presented. The signal selection requires a pair of $b$-tagged jets consistent with those from a Higgs boson decay, and either an electron or a muon from the $W$ boson decay, together with missing transverse momentum from the corresponding neutrino and the stable neutralinos. The analysis is based on data corresponding to 139~$\mathrm{fb}^{-1}$ of $\sqrt{s}=13$ TeV $pp$ collisions provided by the Large Hadron Collider and recorded by the ATLAS detector. No statistically significant evidence of an excess of events above the Standard Model expectation is found.
Limits are set on the direct production of the electroweakinos in simplified models, assuming pure wino cross-sections. Masses of $\tilde{\chi}^{\pm}_{1}/\tilde{\chi}^{0}_{2}$ up to 740 GeV are excluded at 95\% confidence level for a massless $\tilde{\chi}^{0}_{1}$.}
 
\author{The ATLAS Collaboration}
 
\AtlasRefCode{SUSY-2019-08}
 
\AtlasDate{19 September 2019}
 
\arXivId{1909.09226}
 
\AtlasJournalRef{Eur.\ Phys.\ J. C \textbf{80} (2020) 691}
\AtlasDOI{10.1140/epjc/s10052-020-8050-3}

\hypersetup{pdftitle={ATLAS document},pdfauthor={The ATLAS Collaboration}}
 
\begin{document}
 
\maketitle

\section{Introduction}
% The next lines are included from the .//section/intro.tex input file
The Standard Model (SM) is a remarkably successful theory, yet it is clear that this theory is not a complete description of nature. The discovery in 2012 of the SM Higgs boson~\cite{HIGG-2012-27,CMS-HIG-12-028,HIGG-2014-14,HIGG-2015-07}, by the ATLAS and CMS collaborations, confirmed the mechanism of the electroweak symmetry breaking and highlighted the hierarchy problem~\cite{Sakai:1981gr,Dimopoulos:1981yj,Ibanez:1981yh,Dimopoulos:1981zb}.
Supersymmetry (SUSY) \cite{Golfand:1971iw,Volkov:1973ix,Wess:1974tw,Wess:1974jb,Ferrara:1974pu,Salam:1974ig}, a theoretical extension to the SM, resolves the hierarchy problem by introducing a new fermion (boson) supersymmetric partner for each boson (fermion) in the SM.
In SUSY models that conserve \emph{R}-parity~\cite{Farrar:1978xj}, the SUSY particles are produced in pairs. Furthermore, the lightest supersymmetric particle (LSP) is stable and weakly interacting, thus constituting a viable dark-matter candidate~\cite{Goldberg:1983nd,Ellis:1983ew}.
 
In SUSY scenarios the partners of the SM Higgs boson (h) and the gauge bosons, known as the higgsinos, winos (partners of the $\mathrm{SU(2)_L}$ gauge fields), and bino (partner of the $\mathrm{U(1)}$ gauge field) are collectively referred to as electroweakinos. Charginos $\tilde{\chi}_{i}^{\pm}$ $(i=1,2)$ and neutralinos $\tilde{\chi}_{j}^{0}$ $(j=1,2,3,4)$ are the electroweakino mass eigenstates which are linear superpositions of higgsinos, winos, and bino.
For the models considered in this paper, the lightest neutralino ($\tilde{\chi}_{1}^{0}$) is a bino-like LSP. The lightest chargino ($\tilde{\chi}_{1}^{\pm}$) and next-to-lightest neutralino ($\tilde{\chi}_{2}^{0}$) are wino-like and nearly mass degenerate.
 
Naturalness considerations~\cite{Barbieri:1987fn,deCarlos:1993yy} suggest that the lightest of the electroweakinos have masses near the electroweak scale.
In scenarios where the strongly produced SUSY particles are heavier than a few \TeV, the direct production of electroweakinos may be the dominant SUSY production mechanism at the Large Hadron Collider (LHC).
The lightest chargino and next-to-lightest neutralino can decay via $\chinoonepm \rightarrow W \ninoone$ and $\ninotwo \rightarrow h / Z \ninoone$ respectively~\cite{Fayet:1976et,Fayet:1977yc,Gunion:1987} in scenarios where the lepton superpartners are heavier than the electroweakinos. In this case  the decay via the Higgs boson is dominant for many choices of SUSY parameters, as long as $m(\ninotwo)-m(\ninoone)>m(h)$. Scenarios with light electroweakinos also provide a possible explanation for the discrepancy between the muon anomalous magnetic moment $g-2$ measurement and the SM predictions~\cite{PhysRevD.53.6565,PhysRevD.61.095004,PhysRevD.61.095004}.
 
\begin{figure}[h!]

\centering\includegraphics[width=0.48\textwidth]{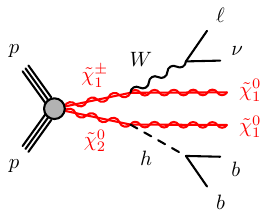}
 
\caption{
A diagram illustrating the signal scenario considered for the production of a chargino and a next-to-lightest neutralino.}
\label{fig:diagrams}
\end{figure}

This paper presents a search for direct production of electroweakinos in proton--proton ($pp$) collisions produced at the LHC at $\sqrt{s} = 13$~\TeV. This analysis is designed to be sensitive to direct production of a chargino and a neutralino that promptly decay as $\chinoonepm \rightarrow W \ninoone$ and $\ninotwo \rightarrow h \ninoone$. The search targets a $W$ boson which decays into an electron or muon (and corresponding neutrino) and a Higgs boson which decays into a pair of $b$-quarks, as shown in Figure~\ref{fig:diagrams}.
The signature consists of exactly one light lepton ($e$ or $\mu$), two jets originating from the fragmentation of $b$-quarks, and missing transverse momentum (\mpt) from neutralinos and neutrinos. A set of simplified SUSY models
is used to optimise the search and interpret the results.
The branching ratios of $\chinoonepm \rightarrow W \ninoone$ and $\ninotwo \rightarrow h \ninoone$ are assumed to be 100\%.
The branching ratio of $h \rightarrow b\bar{b}$ is taken to be 58.3\% as expected for the SM Higgs boson.
 
Previous searches for charginos and neutralinos at the LHC targeting decays via the Higgs boson have been reported by the ATLAS~\cite{SUSY-2017-01} and CMS~\cite{Sirunyan:2018ubx} collaborations. Because of increased integrated luminosity and an improved two-dimensional fit procedure, the search presented here significantly extends the SUSY parameter space sensitivity beyond that of the previously published $13$~\TeV ATLAS search~\cite{SUSY-2017-01} for the same final state.
 
% End of text imported from the .//section/intro.tex input file

\section{ATLAS detector}
% The next lines are included from the .//section/detector.tex input file
The ATLAS detector~\cite{PERF-2007-01} is a multipurpose particle detector with a nearly 4$\pi$ coverage in solid angle.\footnote{ATLAS uses a right-handed coordinate system with its origin at the nominal interaction point in the centre of the detector. The positive $x$-axis is defined by the direction from the interaction point to the centre of the LHC ring, with the positive $y$-axis pointing upwards, while the beam direction defines the $z$-axis. Cylindrical coordinates $(r,\phi)$ are used in the transverse plane, $\phi$ being the azimuthal angle around the $z$-axis. The pseudorapidity $\eta$ is defined in terms of the polar angle $\theta$ by $\eta=-\ln\tan(\theta/2)$. Rapidity is defined as $y$ = $0.5 \ln[(E + p_z)/(E-p_z)]$ where $E$ denotes the energy and $p_z$ is the component of the momentum along the beam direction. The angular distance $\Delta R$ is defined as $\sqrt{(\Delta y)^2 + (\Delta \phi)^2}$.}  It consists of an inner tracking detector surrounded by a thin superconducting solenoid providing a \SI{2} {\tesla} axial magnetic field, electromagnetic and hadron calorimeters, and a muon spectrometer. The inner tracking detector covers the pseudorapidity range \(|\eta| < 2.5\). It consists of silicon pixel, silicon microstrip, and transition radiation tracking detectors. A new inner pixel layer, the insertable B-layer~\cite{ATLAS-TDR-19, PIX-2018-001}, was added at a mean radius of 3.3~cm before the start of 2015 data taking period, improving the identification of $b$-jets. Lead/liquid-argon (LAr) sampling calorimeters provide electromagnetic (EM) energy measurements
with high granularity. A steel/scintillator-tile hadron calorimeter covers the central pseudorapidity range (\(|\eta| < 1.7\)).
The endcap and forward regions are instrumented with LAr calorimeters
for EM and hadronic energy measurements up to \(|\eta| = 4.9\).
The muon spectrometer surrounds the calorimeters and is based on
three large air-core toroidal superconducting magnets with eight coils each.
The field integral of the toroids ranges between 2.0 and 6.0~Tm across most of the detector.
The muon spectrometer includes a system of precision tracking chambers and fast detectors for triggering.
A two-level trigger system \cite{TRIG-2016-01}~is used to select events.
The first-level trigger is implemented in hardware and uses a subset of the detector information to keep the accepted rate below \SI{100}{\kHz}.
This is followed by a software-based trigger that
reduces the accepted event rate to \SI{1}{\kHz} on average
depending on the data-taking conditions.
 
% End of text imported from the .//section/detector.tex input file

\section{Dataset and simulated events}
% The next lines are included from the .//section/datasamples.tex input file
The results were obtained using 139~$\mathrm{fb}^{-1}$ of $pp$ LHC collision data collected between 2015 and 2018 by the ATLAS detector, with a centre-of-mass energy of 13~\TeV\ and a 25~ns proton bunch crossing interval. 
In 2015--2016 the average number of interactions per bunch crossing (pile-up) was $\langle\mu\rangle=20$, increasing to $\langle\mu\rangle=38$ in 2017 and to $\langle\mu\rangle=37$ in 2018.
The uncertainty in the combined 2015--2018 integrated luminosity is 1.7\% \cite{ATLAS-CONF-2019-021}, obtained using the LUCID-2 detector \cite{LUCID2} for the primary luminosity measurements.

Monte Carlo (MC) simulated datasets are used to model the SM backgrounds and evaluate signal selection efficiency and yields. All simulated samples were produced using the ATLAS simulation infrastructure~\cite{SOFT-2010-01} and \textsc{Geant~4}~\cite{Agostinelli:2002hh}, or a faster simulation based on a parameterisation  of the calorimeter response and \textsc{Geant~4} for the other detector systems. All simulated events were generated with a varying number of inelastic pp interactions overlaid on the hard-scattering event to model the multiple proton--proton interactions in the same and nearby bunch crossings. The pile-up events are generated with \pythia~8.186~\cite{Sjostrand:2007gs} using the NNPDF2.3LO set of PDFs~\cite{Ball:2012cx} and the A3 tune~\cite{ATL-PHYS-PUB-2016-017}. The simulated events were reconstructed with the same algorithms as those used for data.

The backgrounds considered in this analysis are: $t\bar{t}$ pair production; single-top production ($s$-channel, $t$-channel, and associated $Wt$ production); $W/Z$+jets production; $t\bar{t}$ production with an electroweak boson ($ttV$); Higgs boson production ($tth$, $Vh$); and diboson ($WW$, $WZ$, $ZZ$) and triboson ($VVV$ where $V=W,Z$) production. Background samples were simulated using different MC event generators depending on the process.
All background processes were normalised to the best available theoretical calculation of their respective cross-sections.
The \SHERPA samples used for $W$+jets modelling include up to two partons at NLO and four partons at LO using Comix~\cite{Gleisberg:2008fv} and OpenLoops~\cite{Cascioli:2011va,Denner:2016kdg} and merged with the \SHERPA parton shower~\cite{Schumann:2007mg} according to the ME+PS@NLO prescription~\cite{Hoeche:2011fd,Hoeche:2012yf,Catani:2001cc,Hoeche:2009rj} using the set of tuned parameters developed by the \SHERPA authors.
The event generators, the parton shower and hadronisation routines, and the underlying-event parameter tunes and parton distribution function (PDF) sets used in simulating the SM background processes, along with the accuracy of the theoretical cross-sections, are all summarised in Table~\ref{tab:MCgen}.
 
For all samples showered with \pythia, the EvtGen~v1.2.0~\cite{Lange:2001uf} program was used to simulate the properties of the bottom- and charm-hadron decays.
Several samples produced without detector simulation were employed to estimate systematic uncertainties associated with the specific configuration of the MC generators used for the nominal SM background samples. They include variations of the renormalisation and factorisation scales, the CKKW-L~\cite{Lonnblad:2012ix} matching scale, as well as different PDF sets and fragmentation/hadronisation models. Details of the MC modelling uncertainties are discussed in Section~\ref{sec:syst}.

The SUSY signal samples were generated using \madgraph~v2.6.2~\cite{Alwall:2014hca}  and \pythia~8.230 with the A14~\cite{ATL-PHYS-PUB-2014-021} set of tuned parameters for the modelling of the parton showering (PS), hadronisation and underlying event. The matrix element (ME) calculation is performed at tree level and include the emission of up to two additional partons. The ME--PS matching is done using the CKKW-L prescription, with a matching scale set to one quarter of the chargino and next-to-lightest neutralino mass. The NNPDF2.3LO~\cite{Ball:2012cx} PDF set was used.

Signal cross-sections are calculated at next-to-leading-order (NLO) accuracy in the strong coupling constant, adding the resummation of soft gluon emission at next-to-leading-logarithm accuracy (NLO+NLL)~\cite{Debove:2010kf,Fuks:2012qx,Fuks:2013vua,Fiaschi:2018hgm}. The nominal cross-section and its uncertainty are taken as the midpoint and half-width of an envelope of cross-section predictions using different PDF sets and factorisation and renormalisation scales, as described in Ref.~\cite{Borschensky:2014cia}.
The simplified model has two parameters, the first being the mass of the $\tilde\chi^\pm_1$ and $\tilde\chi^0_2$  (which are assumed to be equal), and the second being the mass of the $\tilde\chi^0_1$.
The signal cross-sections decrease as the $\tilde\chi^\pm_1/\tilde\chi^0_2$ mass increases, ranging from 769~fb for a 250~\GeV\ $\tilde\chi^\pm_1/\tilde\chi^0_2$ mass to 1.3~fb for a 1000~\GeV\ $\tilde\chi^\pm_1/\tilde\chi^0_2$ mass.

\begin{table}
\begin{center}
\footnotesize
\resizebox{\textwidth}{!}{
\begin{tabular} {l | c  c  c  c  c }
\hline 
Process & Generator & Parton shower and        & Tune & PDF & Cross-section \\
&           & hadronisation &      &     &               \\
\hline 
$t\bar{t}$  & \POWHEGBOX v2~\cite{Alioli:2010xd,Frixione:2007vw,Frixione:2007nw,Nason:2004rx}    & \pythia 8.230~\cite{Sjostrand:2007gs}   & A14~\cite{ATL-PHYS-PUB-2014-021}     & NNPDF2.3LO~\cite{Ball:2012cx}   & NNLO+NNLL~\cite{Czakon:2011xx}\\
 
Single top  & \POWHEGBOX v2~\cite{Re:2010bp,Frederix:2012dh,Alioli:2009je}    & \pythia 8.230   & A14     & NNPDF2.3LO   & NLO+NNLL~\cite{Aliev:2010zk}\\
 
$W/Z$+jets  & \SHERPA 2.2.1~\cite{Gleisberg:2008ta}    & \SHERPA 2.2.1   & \SHERPA standard & NNPDF3.0NNLO  & NNLO~\cite{Anastasiou:2003ds} \\
 
Diboson     & \SHERPA 2.2.1 \& 2.2.2     & \SHERPA 2.2.1 \& 2.2.2   & \SHERPA standard & NNPDF3.0NNLO  & NLO\\
 
Triboson    & \SHERPA 2.2.1 \& 2.2.2   & \SHERPA 2.2.1 \& 2.2.2   & \SHERPA standard & NNPDF3.0NNLO  & NLO\\
 
$t\bar{t}+V$&\madgraph~v2.3.3 & \pythia 8.210    & A14     & NNPDF2.3LO   &NLO~\cite{Baernreuther:2012ws} \\
 
$tth$       & \POWHEGBOX v2   & \pythia 8.230    & AZNLO~\cite{STDM-2012-23}     & CTEQ6L1~\cite{Nadolsky:2008zw} & NLO~\cite{deFlorian:2016spz}\\
 
$Vh$        & \POWHEGBOX v2   & \pythia 8.212    &  A14     & NNPDF2.3LO   &NLO~\cite{deFlorian:2016spz}  \\
 
\hline 
\end{tabular}
}
\caption{Overview of MC generators used for different simulated event samples.}
\label{tab:MCgen}
\end{center}
\end{table}
% End of text imported from the .//section/datasamples.tex input file

\section{Event reconstruction}
% The next lines are included from the .//section/reco.tex input file
 
Events are required to have at least one reconstructed interaction vertex with a minimum of two associated tracks each having $\pt>500$~\MeV. In events with multiple vertices, the one with the highest sum of squared transverse momenta of associated tracks is chosen as the primary vertex (PV)~\cite{ATL-PHYS-PUB-2015-026}. A set of baseline quality criteria are applied to reject events with non-collision backgrounds or detector noise~\cite{ATLAS-CONF-2015-029}.
 
Two identification levels are defined for leptons and jets: `baseline' and `signal'.
Baseline leptons and jets are selected with looser identification criteria, and are used in computing the missing transverse momentum as well as in resolving possible reconstruction ambiguities.
Signal leptons and jets are a subset of the baseline objects with tighter quality requirements which are used to define the search regions. Isolation criteria, defined with a list of tracking-based and calorimeter-based variables, are used to select signal leptons by discriminating against semileptonic heavy-flavour decays and jets misidentified as leptons.
 
Electron candidates are reconstructed from energy deposits in the electromagnetic calorimeter that are matched to charged-particle tracks in the inner detector (ID)~\cite{Aad:2019tso}.
Baseline electrons are required to satisfy $\pt>7$~\GeV\ and $|\eta|<2.47$. They are identified using the `loose' operating point provided by a likelihood-based algorithm, described in Ref.~\cite{Aad:2019tso}.
The number of hits in the innermost pixel layer is used to discriminate between electrons and converted photons. The longitudinal impact parameter $z_0$ relative to the PV is required to satisfy $|z_0\sin\theta| <0.5$ mm.
The `tight' likelihood operating point is applied for signal electron identification and the significance of the transverse impact parameter $d_0$ must satisfy $|d_0/\sigma(d_0)|<5$.
Signal electron candidates with $\pt < 200$~\GeV\ are further refined using the \emph{FCLoose} isolation working point, while those with larger $\pt$ are required to pass the \emph{FCHighPtCaloOnly} isolation working point, as described in Ref.~\cite{Aad:2019tso}.
 
Muon candidates are reconstructed from matching tracks in the ID and muon spectrometer, refined through a global fit which uses the hits from both subdetectors~\cite{PERF-2015-10}.
Baseline muons must have $\pt>6$~\GeV\ and $|\eta|<2.7$, and satisfy the `medium' identification criteria.
Similarly to electrons, the longitudinal impact parameter $z_0$ relative to the PV is required to satisfy $|z_0\sin\theta| <0.5$~mm.
Signal muon candidates are further defined with tighter pseudorapidity and impact parameter requirements, $|\eta|<2.5$ and $|d_0/\sigma(d_0)|<3$.
The \emph{FCLoose} isolation working point is also required for signal muons~\cite{PERF-2015-10}.
 
Jets are reconstructed from three-dimensional topological energy clusters in the calorimeters using the anti-$k_t$ algorithm \cite{PERF-2014-07}~with a radius parameter $R=0.4$~\cite{Cacciari:2008gp}.
Baseline jets are selected in the region $|\eta|<4.5$ and have $\pt>20$~\GeV.
To suppress jets from pile-up interactions, the jets with $|\eta|<2.8$ and $\pt<120$~\GeV\ are required to satisfy the `medium' working point of the jet vertex tagger (JVT),  a tagging algorithm that identifies jets originating from the PV using track information~\cite{PERF-2014-03, FastJet}.
The selection of signal jets is further refined by requiring them to be in the region $|\eta|<2.8$ and have $\pt>30$~\GeV.
 
Jets containing $b$-hadrons are identified as `$b$-tagged' using the MV2c10 algorithm, a multivariate discriminant based on the track impact parameters and displaced secondary vertices~\cite{FTAG-2018-01}. These
$b$-tagged jets are reconstructed in the region $|\eta|<2.5$ and have $\pt>30$~\GeV.
The $b$-tagging working point provides an efficiency of 77\% for jets containing $b$-hadrons in simulated \ttbar events, with rejection rates of 110 and 4.9 for light-flavour jets and jets containing $c$-hadrons, respectively~\cite{ATL-PHYS-PUB-2017-013}.
 
To resolve the reconstruction ambiguities between electrons, muons and jets, an overlap removal procedure is applied to baseline objects.
First, any electron sharing the same ID track with a muon is rejected. If it shares the same ID track with another electron, the one with lower \pt is discarded.
Next, jets are rejected if they lie within  $\Delta R = 0.2$ of a muon or if the muon is matched to the jet through ghost association~\cite{ghostassociation:2008gn}.
Subsequently, electrons within a cone of size $\Delta R = \min(0.4,0.04+10$ \GeV$/\pt)$ around a jet are removed.
Last, muons within a cone, defined in the same way as for electrons, around any remaining jet are removed.

The missing transverse momentum \mpt, with magnitude \met\, is calculated as the negative vectorial sum of the transverse momentum of all baseline reconstructed objects (electrons, muons, jets and photons~\cite{PERF-2017-02}) and the soft term. The soft term includes all tracks associated with the PV but not matched to any reconstructed physics object. Tracks not associated with the PV are not considered in the \met calculation, improving the \met resolution by suppressing the effect of pile-up~\cite{PERF-2016-07,ATLAS-CONF-2018-023}.

Corrections are applied to simulated events in order to account for the trigger, particle identification, and reconstruction efficiency differences between data and simulation.
 
% End of text imported from the .//section/reco.tex input file

\section{Event selection}
% The next lines are included from the .//section/eventselection.tex input file

Events are recorded with the lowest-threshold
\met trigger available, which is fully efficient for selecting events when the offline requirement of $\met > 240$~\GeV\ is applied. To target the signal events, which have a leptonically decaying $W$ boson and a Higgs boson decaying into a $b\bar{b}$ pair, events are required to have exactly one signal electron or muon (but not both) and either two or three signal jets,  two of which must be $b$-tagged.
The signal regions (SR) are defined using variables which suppress background contributions and increase the sensitivity for signal. These variables are based on the kinematic properties of the $b$-jets, the lepton and the missing transverse momentum, and are defined as follows:
 
\begin{itemize}

\item The invariant mass of the two $b$-jets, \mbb, is required to be in the range $100 < \mbb < 140$~\GeV, in order to preferentially select $b$-jets from the Higgs boson decays.
 
\item The invariant mass of the lepton and the leading $b$-jet is denoted by $m(\ell,b_1)$. For \ttbar\ or single-top (particularly the $Wt$-channel) backgrounds, if the lepton and the leading $b$-jet originate from the same top-quark, the $m(\ell,b_1)$ distribution has an endpoint at $\sqrt{m^{2}(t)-m^{2}(W)}$. For signal events, the lepton and $b$-jet are produced from the $\tilde{\chi}_{1}^{\pm}$ and $\tilde{\chi}_{2}^{0}$ decay chains, respectively. The distribution of the invariant mass depends on the mass of the SUSY particles. For signal events with high-mass $\tilde{\chi}_{1}^{\pm}$/$\tilde{\chi}_{2}^{0}$, this observable provides good discrimination against background events.
 
\item The transverse mass, \mt, is defined from the lepton transverse momentum ${\boldsymbol p}_\mathrm{T}^{\ell}$ and \mpt\ as
\begin{linenomath}
\begin{equation}
m_{\mathrm{T}} =
\sqrt{2 p_{\mathrm{T}}^{\ell} \met
(1-\cos[\Delta\phi({\boldsymbol p}_{\mathrm{T}}^{\ell},\mpt)])},
\nonumber
\end{equation}
\end{linenomath}
where $\Delta\phi({\boldsymbol p}_{\mathrm{T}}^{\ell},\mpt)$ is the azimuthal angle between ${\boldsymbol p}_{\mathrm{T}}^{\ell}$ and \mpt. For $W$+jets and semileptonic \ttbar\ events, in which one on-shell $W$ boson decays leptonically, the observable has an upper endpoint at the $W$ boson mass. The \mt\ distribution for signal events extends significantly above $m(W)$.

\item The contransverse mass~\cite{Tovey:2008ui, Polesello:2009rn} of two $b$-jets, \mctbb, is defined as:
\begin{linenomath}
\begin{equation}
\mctbb =  \sqrt{2 \pt^{b_1} \pt^{b_2}  \left ( 1+\cos\Delta\phi_{bb} \right )}, \notag
\label{eq:mctbb}
\end{equation}
\end{linenomath}
where $\pt^{b_1}$ and $\pt^{b_2}$ are the transverse momenta of the two leading $b$-jets and $\Delta\phi_{bb}$ is the azimuthal angle between them.
For the \ttbar\ background, the observable has an upper endpoint at $(m^{2}(t)-m^{2}(W))/m(t)$.
A requirement that \mctbb\ be larger than 180~\GeV\ efficiently suppresses the \ttbar\ background.
 
\end{itemize}

An overview of the signal region definitions is provided in Table~\ref{tab:SignalRegionDef}.
Three separate classes of signal regions are defined, progressively targeting increasing mass differences between the $\chinoonepm$ (and its mass-degenerate $\ninotwo$ wino partner) and the $\ninoone$.
These regions are labelled SR-LM, SR-MM and SR-HM to indicate low (LM), medium (MM) and high (HM) mass differences respectively.
Requirements on \mt\ make the three regions mutually exclusive.
Of the three signal regions, SR-LM selects the smallest values of \mt. It targets signal models with a mass-splitting between the $\ninotwo$ (and hence the $\chinoonepm$) and the $\ninoone$ that is similar to the Higgs boson mass.
The other two signal regions select progressively larger mass differences by requiring larger values of \mt.
The signal region with the highest requirement on \mt, SR-HM, also requires $m(\ell,b_1) > 120$~\GeV\ in order to further suppress \ttbar\ and single-top background events.
The three signal regions otherwise share a common set of selections on \met, \mbb and \mct.

When setting model-dependent exclusion limits (`excl.'), each of the three SRs is binned in three \mctbb regions, thus providing nine bins in total for a simultaneous two-dimensional fit in \mctbb\ and \mt\ across the three SRs. This multi-bin approach enhances the sensitivity to a range of SUSY scenarios with different properties.
For model-independent limits and null-hypothesis tests (`disc.' for discovery), the various \mctbb bins are merged for each of the three SRs. The requirement of $m(\ell,b_1) > 120$~\GeV~is only applied in SR-HM. Furthermore, the upper bound on \mt is removed for SR-LM and SR-MM. The fit strategy is detailed in Section~\ref{sec:bkg}. The systematic uncertainties, fit and results discussed in the following sections are based on the exclusion SRs, while the model-independent results are based on the discovery SRs.

\begin{table}
\begin{center}
\begin{tabular} {l | c c c }
\hline 
&  \textbf{SR-LM} & \textbf{SR-MM} & \textbf{SR-HM} \\
\hline
$N_{\mathrm{lepton}}$ & \multicolumn{3}{c}{$=$ 1}\\
$\ptl$ [\GeV] & \multicolumn{3}{c}{ $>7(6)$ for $e$($\mu$)} \\
$N_\mathrm{jet}$ & \multicolumn{3}{c}{$=$ 2 or 3}\\
$N_{b\textrm{-jet}}$ &\multicolumn{3}{c}{$=$ 2} \\
\met $[\SI{}{\GeV}]$ & \multicolumn{3}{c}{$>240$}\\
\mbb  $[\SI{}{\GeV}]$ & \multicolumn{3}{c}{$\in [100,140]$}\\
$m(\ell,b_1)$ $[\SI{}{\GeV}]$ & -- & -- & $>120$ \\
\hline
\mt $[\SI{}{\GeV}]~\mathrm{(excl.)}$&   $\in [100,160]$ & $\in [160,240]$ & $>240$ \\

\mct $[\SI{}{\GeV}]~\mathrm{(excl.)}$ &\multicolumn{3}{c}{ $ \{ \in [180,230]$,\,$\in [230,280]$, $>280  \}$}\\
 
\hline
\mt $[\SI{}{\GeV}]~\mathrm{(disc.)}$&   $>100$ & $>160$ & $>240$ \\
\mct $[\SI{}{\GeV}]~\mathrm{(disc.)}$ & \multicolumn{3}{c}{ $>180$}\\
\hline
\hline 
\end{tabular}
\caption{Overview of the selection criteria for the signal regions. Each of the three `excl.' SRs is binned in three \mct\ regions for a total of nine `excl.'  bins.}
\label{tab:SignalRegionDef}
\end{center}
\end{table}
% End of text imported from the .//section/eventselection.tex input file

\section{Background estimation}
\label{sec:bkg}
% The next lines are included from the .//section/background.tex input file
The expected backgrounds in each signal region are determined in a profile likelihood fit, referred to as a `background-only fit'. In this fit, the normalisation of the backgrounds is adjusted to match the data in control regions with negligible signal contamination.
The resulting normalisation factors are then used to correct the expected yields of the corresponding backgrounds in the various signal regions. The control regions -- as detailed in the following -- are designed to be enriched in the major background processes: \ttbar, single-top and $W$+jets processes.
The control region for single-top has a similar composition in single-top processes as the signal regions, therefore a single scale factor is used.
All CRs are designed to be non-overlapping with the signal regions and also mutually exclusive. A probability density function is defined for each of the control regions. The inputs are the observed event yield and the predicted background yield from simulation with Poisson statistical uncertainties as well as with systematic uncertainties as nuisance parameters. The nuisance parameters are constrained by Gaussian distributions with widths corresponding to the sizes of the uncertainties.
The uncertainties do not only vary the scales, but also account for bin-to-bin transitions.
The systematic uncertainties are detailed in Section~\ref{sec:syst}. The product of all the probability density functions forms the likelihood. Normalisation and nuisance parameters are correlated in all regions participating in the fit. The likelihood is maximised by adjusting the normalisation and nuisance parameters. The extrapolation of the adjusted normalisation and nuisance parameters to the signal regions is  checked in validation regions (VR), as defined below, which kinematically resemble the signal regions but are expected to have less signal. The VRs do not overlap with the CRs or SRs.
 
Subdominant background processes, such as $Z$+jets, diboson and multiboson, \ttbar+$V$, \ttbar+$h$ and $Vh$, which have no dedicated control regions, are normalised to the cross-sections indicated in Table~\ref{tab:MCgen}. In the same way as for the dominant backgrounds, their expected yields in the SRs are subject to statistical and systematic uncertainties. Backgrounds with fake leptons such as jets misreconstructed as a lepton, and events with leptons originating from a jet produced by heavy-flavour quarks or from photon conversions are estimated using a matrix method as described in Ref.~\cite{MM:2014}, and found to be negligible in all regions.
 
The $t\bar{t}$ background estimation relies on a set of three CRs (labelled TR-LM, TR-MM, TR-HM), each with an \mt\ selection the same as in the SRs. In order to obtain samples enriched in $t\bar{t}$ events, the requirement on $m(\ell,b_1)$ is removed and the selection criteria for \mct\ and \mbb\ are inverted relative to the SRs. These three control regions are fit simultaneously to obtain a single normalisation factor. The $W$+jets contributions in the SRs are constrained by a single CR (labelled WR) defined similarly to the SRs but with less stringent lower bounds on \mt and an off-peak region for \mbb.
The fraction of events with heavy flavor hadrons in simulated $W$+jet events was found similar between the WR and the SRs. Events in the single-top CR (labelled STR) must satisfy the SR requirements except that this CR requires $\mt > 100$~\GeV\ and $\mbb > 195$~\GeV.
The $t\bar{t}$ purity varies from ~79\% in TR-LM to ~86\% in TR-MM. The purity of the single-top ($W$+jets) is ~52\% (~53\%) in STR (WR).
 
Two sets of VRs are defined for each SR, including the off-peak (\mbb $<$100 or $>$140 GeV) and the on-peak \mbb\ regions, with the same \mt\ as in the SR.
The on-peak VRs validate the extrapolation from the CRs to the SRs in \mbb, and the off-peak VRs validate the extrapolation in \mct.
The validation regions share the same \mt binning as the signal regions, denoted as LM, MM and HM. The background modelings are validated in low, medium and high \mt regions separately.
A summary of all CR and VR selection criteria is reported in Table~\ref{tab:CRVRdef}.
 
\begin{table}
\begin{center}
\resizebox{\textwidth}{!}{
\begin{tabular} {l | c c c | cccccc}
\hline 
\textbf{CR} & \textbf{TR-LM} &  \textbf{TR-MM} &  \textbf{TR-HM} &  \multicolumn{3}{c}{\textbf{WR}} & \multicolumn{3}{c}{ \textbf{STR}} \\
\hline 
\mbb [\GeV]  & \multicolumn{3}{c|}{$<$100 or $>$140} & \multicolumn{3}{c}{$\in[50,80]$} &\multicolumn{3}{c}{$>$195}\\
\mt [\GeV]   & $\in[100,160]$ & $\in[160,240]$ & $>$240 & \multicolumn{3}{c}{$\in[50,100]$} &\multicolumn{3}{c}{$>$100}\\
\mct [\GeV]  &\multicolumn{3}{c|}{$<$180} & \multicolumn{3}{c}{$>$180} &\multicolumn{3}{c}{$>$180}\\
\hline 
\textbf{VR} & \textbf{VR-onLM} &  \textbf{VR-onMM} & \textbf{VR-onHM} & \multicolumn{2}{c}{ \textbf{VR-offLM}} &  \multicolumn{2}{c}{\textbf{VR-offMM}} &  \multicolumn{2}{c}{\textbf{VR-offHM}}\\
\hline 
\mbb [\GeV]  & \multicolumn{3}{c|}{$\in[100,140]$} & \multicolumn{2}{c}{$\in[50,80]\, \cup [160, 195]$} & \multicolumn{2}{c}{$\in[50,80]\, \cup [160, 195]$}& \multicolumn{2}{c}{$\in[50,75]\, \cup [165, 195]$}\\
\mt [\GeV]   & $\in[100,160]$ & $\in[160,240]$ & $>$240 & \multicolumn{2}{c}{$\in[100,160]$} & \multicolumn{2}{c}{$\in[160,240]$} & \multicolumn{2}{c}{$>$240}\\
\mct [\GeV]  &\multicolumn{3}{c|}{$<$180}&\multicolumn{6}{c}{$>$180}\\
\hline 
\hline 
\end{tabular}}
\caption{Overview of the CR and VR definitions. All regions partially share the same selection as the SR for all variables except $m(\ell,b_1)$, which is not used in the CR and VR definitions.}
\label{tab:CRVRdef}
\end{center}
\end{table}
 
% End of text imported from the .//section/background.tex input file

\section{Systematic uncertainties}
\label{sec:syst}
% The next lines are included from the .//section/syst.tex input file
Systematic uncertainties are evaluated for all simulated signal and background events.
For the dominant backgrounds with dedicated control regions, the systematic uncertainties impact the extrapolation from
the control regions to the corresponding signal regions.
For all other backgrounds estimated from simulation, the uncertainties affect the overall cross-section normalisation and the acceptance of the analysis selection.
Uncertainties arising from theoretical modelling and detector effects are estimated and discussed below.
A breakdown of the dominant systematic uncertainties in background estimates in the various exclusion signal regions is summarised in Table~\ref{systematicssummary}.
The uncertainties in the scale factor fits to the control regions are listed as `Normalisation of dominant backgrounds'.
 
Several uncertainties in the theoretical modelling of the single-top and \ttbar backgrounds are considered. Uncertainties due to the choice of hard-scatter generation program are
estimated by comparing \POWHEGBOX generated events, showered using \PYTHIA 8, with events generated by a\textsc{MC@NLO} and showered with \PYTHIA 8, while those due to the choice of parton shower model are evaluated by comparing \POWHEGBOX generated samples showered using \PYTHIA 8 with \POWHEGBOX samples showered using \HERWIG 7\cite{Bellm:2015jjp}.
The uncertainties from the modelling of initial- and final-state radiation are assessed by varying the renormalisation and factorisation scales up and down by a factor of two, with the radiation setting varied as well~\cite{ATL-PHYS-PUB-2016-004}.
For single-top $Wt$ production, the impact of interference between single-resonant and double-resonant top-quark production is estimated
by comparing the nominal sample generated using the diagram removal method with a sample generated using the diagram subtraction method.
For the different signal regions, the dominant uncertainty sources are the \ttbar parton shower in SR-LM (10\%), and the single-top generator uncertainties in SR-MM (10\%) and SR-HM (21\%).
 
Theory uncertainties affecting the generator predictions for $W/Z$+jets, diboson, triboson and \ttbar+$W/Z$ samples
are estimated by taking the envelope of the seven-point variations of the renormalisation and factorisation scales.
For $W/Z$+jets, the uncertainties from the PDF variations, as well as from the variations of matching and resummation scales are also considered.
Additionally, an overall 6\% (20\%) systematic uncertainty in the inclusive cross-section is assigned for the diboson (triboson) sample~\cite{ATL-PHYS-PUB-2016-002}.
Similar cross-section uncertainties are also assigned for other small background contributions.
 
Theory uncertainties in the expected yields for SUSY signals are estimated by varying by a factor of two the parameters corresponding to the factorisation, renormalisation, and CKKW-L matching scales, as well as the \pythia~8 shower tune parameters. The overall uncertainties range from about 10\% in the region with a large splitting between the $\tilde{\chi}_{2}^{0}/\tilde{\chi}_{1}^{\pm}$ and  $\tilde{\chi}_{1}^{0}$ masses to about 25\% in the mass spectra with small mass splitting.
 
The dominant detector systematic effects are the uncertainties associated with the jet energy scale (JES) and jet energy resolution (JER), the \met modelling, and pile-up.
The jet uncertainties are measured as a function of the \pt\ and $\eta$ of the jet, the pile-up conditions and the jet flavour composition. They are determined using a combination of data and simulation, through measurements of the jet \pt\ balance in dijet, $Z$+jets and $\gamma$+jets events~\cite{PERF-2016-04}.
The systematic uncertainties in the \met modelling are derived by propagating the uncertainties in the energy and momentum scale of each of the objects entering the calculation, and the uncertainties in the soft term's resolution and scale~\cite{ATLAS-CONF-2018-023}.
A pile-up reweighting procedure is applied to simulation to match the distribution of the number of reconstructed vertices observed in data.
The corresponding uncertainty is derived by a reweighting in which $\langle \mu\rangle$ is varied by $\pm4\%$.
The experimental uncertainties have a less significant impact than the theoretical ones in all signal regions: the largest experimental source contributes 5--10\% depending on the SR. The MC statistical uncertainties contribute 5--18\% depending on the SR.
 
\begin{table}
\begin{center}
\setlength{\tabcolsep}{0.0pc}
\begin{tabular*}{\textwidth}{@{\extracolsep{\fill}}lccc}
\noalign{\smallskip}\hline\noalign{\smallskip}
{\textbf{Signal Region}}                                    & SR-LM            & SR-MM            & SR-HM            \\
\noalign{\smallskip}\hline\noalign{\smallskip}
Total background expectation             &  $27$        &  $8.6$        &  $8.1$       \\
\noalign{\smallskip}\hline\noalign{\smallskip}
Total uncertainty               & $\pm 4\ [15\%]~$        & $\pm 2.2\ [25\%] $        & $~\pm 2.7\ [34\%] $             \\
\noalign{\smallskip}\hline\noalign{\smallskip}
\noalign{\smallskip}\hline\noalign{\smallskip}
\multicolumn{4}{c}{Theoretical systematic uncertainties}\\
\noalign{\smallskip}\hline\noalign{\smallskip}
\ttbar\          & $~~\pm 2.6\ [10\%] $          & $\pm 0.6\ [7\%]~ $          & $~\pm 0.33\ [4\%] $       \\
Single top          & $~~~\pm 0.8\ [2.7\%] $          & $~\pm 1.1\ [12\%] $          & $~\pm 1.9\ [23\%] $       \\
$W$+jets         & $~~~~~\pm 0.23\ [0.9\%] $          & $~~~~\pm 0.07\ [0.8\%] $          & $~~~~\pm 0.19\ [2.3\%] $       \\
Other backgrounds      & $~~~~~\pm 0.13\ [0.5\%] $          & $~~~~\pm 0.15\ [1.7\%] $          & $~~~~\pm 0.08\ [1.0\%] $       \\
\noalign{\smallskip}\hline\noalign{\smallskip}
\multicolumn{4}{c}{MC statistical uncertainties}\\
\noalign{\smallskip}\hline\noalign{\smallskip}
MC statistics         & $\pm 1.7\ [6\%] $          & $~\pm 1.1\ [13\%] $          & $~\pm 1.2\ [14\%] $       \\
\noalign{\smallskip}\hline\noalign{\smallskip}
\multicolumn{4}{c}{Uncertainties in the background normalisation}\\
\noalign{\smallskip}\hline\noalign{\smallskip}
Normalisation of dominant backgrounds         & $\pm 1.3\ [5\%] $          & $~\pm 1.6\ [18\%] $          & $~\pm 1.3\ [16\%] $       \\
\noalign{\smallskip}\hline\noalign{\smallskip}
\multicolumn{4}{c}{Experimental systematic uncertainties}\\
\noalign{\smallskip}\hline\noalign{\smallskip}
\met/JVT/pile-up/trigger         & $\pm 1.8\ [7\%] $          & $\pm 0.4\ [4\%]~ $          & $\pm 0.4\ [5\%]~ $       \\
Jet energy resolution         & $\pm 1.6\ [6\%] $          & $\pm 0.5\ [6\%]~ $          & $\pm 0.4\ [5\%]~ $       \\
$b$-tagging         & $\pm 1.1\ [4\%] $          & $~~~~\pm 0.29\ [3.4\%] $          & $~~~~\pm 0.13\ [1.5\%] $       \\
Jet energy scale         & $~~~\pm 0.9\ [3.2\%] $          & $~\pm 0.9\ [10\%] $          & $~\pm 0.29\ [4\%] $       \\
Lepton uncertainties         & $~~~~~\pm 0.32\ [1.2\%] $          & $~~~~\pm 0.09\ [1.0\%] $          & $~~~~\pm 0.19\ [2.3\%] $       \\
\noalign{\smallskip}\hline\noalign{\smallskip}
\end{tabular*}
\end{center}
\caption[Breakdown of uncertainty in background estimates]{
Breakdown of the dominant systematic uncertainties in background estimates in the various exclusion signal regions.
The individual uncertainties can be correlated, and do not necessarily add up in quadrature to
the total background uncertainty. The percentages show the size of the uncertainty relative to the total expected background.
\label{systematicssummary}}
\end{table}

% End of text imported from the .//section/syst.tex input file

\section{Results}
% The next lines are included from the .//section/result.tex input file
 
The observed event yield in each of the exclusion signal regions is summarised in Table~\ref{table.results.yields.fit.SRLMEM} along with the corresponding Standard Model predictions obtained from the background-only fit.
The background normalisation factors are  $1.02^{+0.07}_{-0.09}$  for \ttbar,  $0.6^{+0.5}_{-0.25}$ for single top, and $1.22^{+0.26}_{-0.24}$ for $W$+jets.

% The next lines are included from the .//tables/results/MyTable1LepSRMerge_bkgonly_sum.tex input file
\begin{table}
\newcolumntype{C}{@{}>{{}}c<{{}}@{}}
\begin{center}
\setlength{\tabcolsep}{0.0pc}
{\small
\begin{tabular*}{\textwidth}{@{\extracolsep{\fill}}l*4{C}}
\noalign{\smallskip}\hline\noalign{\smallskip}
{\textbf{ SR-LM}}           & All \mct bins          & Low \mct         & Medium \mct        & High \mct    \\[-0.05cm]
\noalign{\smallskip}\hline\noalign{\smallskip}
Observed           & $34$              & $16$              & $11$              & $7$                    \\
\noalign{\smallskip}\hline\noalign{\smallskip}
Expected          & $27 \pm 4~~$          & $8.8 \pm 2.8$          & $11.3 \pm 3.1~~$          & $7.3 \pm 1.5$              \\
\noalign{\smallskip}\hline\noalign{\smallskip}
\ttbar\          & $16.2 \pm 3.4~~$          & $4.4 \pm 2.2$          & $7.3 \pm 2.5$          & $4.6 \pm 1.2$              \\
Single top          & $2.7 \pm 1.8$          & $1.3 \pm 1.1$          & $0.9_{-0.9}^{+1.0}~$          & $0.6 \pm 0.6$              \\
$W$+jets           & $5.5 \pm 2.0$          & $2.0 \pm 0.9$          & $2.4 \pm 1.3$          & $1.1 \pm 0.5$              \\
Di-/Multiboson          & $0.67 \pm 0.19$          & $0.39 \pm 0.13$          & $0.09_{-0.09}^{+0.11}~$          & $0.18 \pm 0.04$              \\
Others          & $2.23 \pm 0.29$          & $0.81 \pm 0.25$          & $0.64 \pm 0.15$          & $0.77 \pm 0.12$              \\
\noalign{\smallskip}\hline\noalign{\smallskip}
\noalign{\smallskip}\hline\noalign{\smallskip}
{\textbf{ SR-MM}}           & All \mct bins          & Low \mct         & Medium \mct        & High \mct    \\[-0.05cm]
\noalign{\smallskip}\hline\noalign{\smallskip}
Observed           & $13$              & $4$              & $7$              & $2$                    \\
\noalign{\smallskip}\hline\noalign{\smallskip}
Expected          & $8.6 \pm 2.2$          & $4.6 \pm 1.7$          & $2.6 \pm 1.3$          & $1.4 \pm 0.6$              \\
\noalign{\smallskip}\hline\noalign{\smallskip}
\ttbar\          & $2.7 \pm 1.4$          & $1.6 \pm 0.9$          & $0.8 \pm 0.7$          & $0.30 \pm 0.24$              \\
Single top          & $2.7 \pm 1.9$          & $1.6 \pm 1.5$          & $1.0_{-1.0}^{+1.1}~$          & $0.15_{-0.15}^{+0.19}~$              \\
$W$+jets          & $1.5 \pm 0.7$          & $0.6 \pm 0.4$          & $0.3_{-0.3}^{+0.4}~$          & $0.57 \pm 0.26$              \\
Di-/Multiboson          & $0.29 \pm 0.08$          & $0.09 \pm 0.04$          & $0.065 \pm 0.028$          & $0.14 \pm 0.06$              \\
Others          & $1.33 \pm 0.27$          & $0.69 \pm 0.20$          & $0.40 \pm 0.13$          & $0.24 \pm 0.09$              \\
\noalign{\smallskip}\hline\noalign{\smallskip}
\noalign{\smallskip}\hline\noalign{\smallskip}
{\textbf{ SR-HM}}           & All \mct bins          & Low \mct         & Medium \mct        & High \mct    \\[-0.05cm]
\noalign{\smallskip}\hline\noalign{\smallskip}
Observed           & $14$              & $6$              & $5$              & $3$                    \\
\noalign{\smallskip}\hline\noalign{\smallskip}
Expected          & $8.1 \pm 2.7$          & $4.1 \pm 1.9$          & $2.9 \pm 1.3$          & $1.1 \pm 0.5$              \\
\noalign{\smallskip}\hline\noalign{\smallskip}
\ttbar\          & $1.4 \pm 0.5$          & $0.8 \pm 0.4$          & $0.36 \pm 0.25$          & $0.22 \pm 0.15$              \\
Single top          & $2.0_{-2.0}^{+2.4}~$          & $0.9_{-0.9}^{+1.5}~$         & $0.9 \pm 0.9$          & $0.16_{-0.16}^{+0.26}~$              \\
$W$+jets           & $3.7 \pm 1.0$          & $1.9 \pm 0.8$          & $1.4 \pm 0.8$          & $0.45 \pm 0.19$              \\
Di-/Multiboson          & $0.21 \pm 0.06$          & $0.057 \pm 0.025$          & $0.075 \pm 0.027$          & $0.08 \pm 0.04$              \\
Others          & $0.74 \pm 0.16$          & $0.34 \pm 0.09$          & $0.19 \pm 0.08$          & $0.21 \pm 0.08$              \\
\noalign{\smallskip}\hline\noalign{\smallskip}
 
\end{tabular*}
}
\end{center}
\caption{ Background fit results for the exclusion SR regions. 
The errors shown are the statistical plus systematic uncertainties.
Uncertainties in the fitted yields are symmetric by construction,
except where the negative error is truncated at an event yield of zero.
}
\label{table.results.yields.fit.SRLMEM}
\end{table}
% End of text imported from the .//tables/results/MyTable1LepSRMerge_bkgonly_sum.tex input file
 
\begin{figure}
\begin{center}
\includegraphics[width=0.45\textwidth]{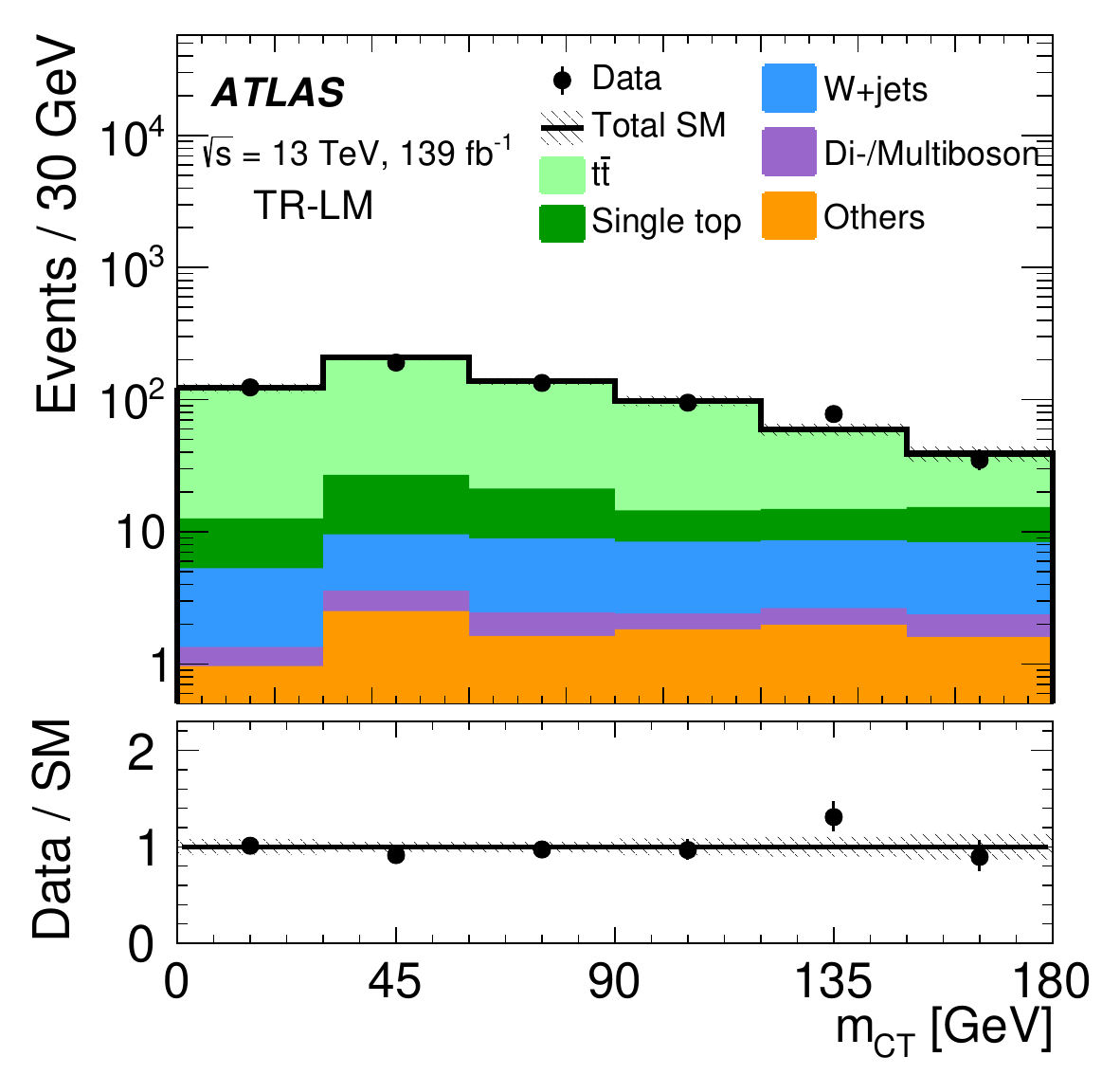}
\includegraphics[width=0.45\textwidth]{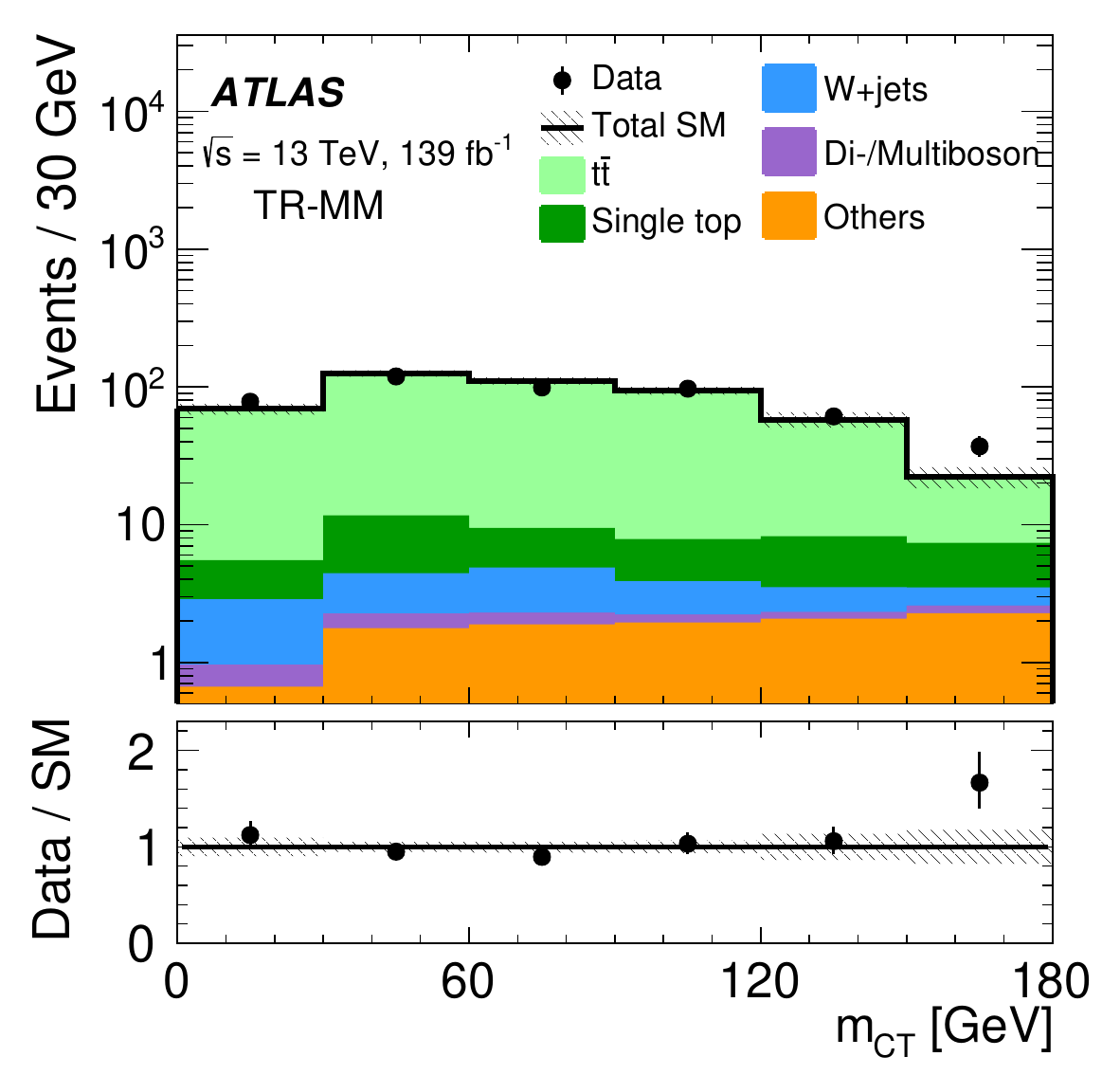}
\includegraphics[width=0.45\textwidth]{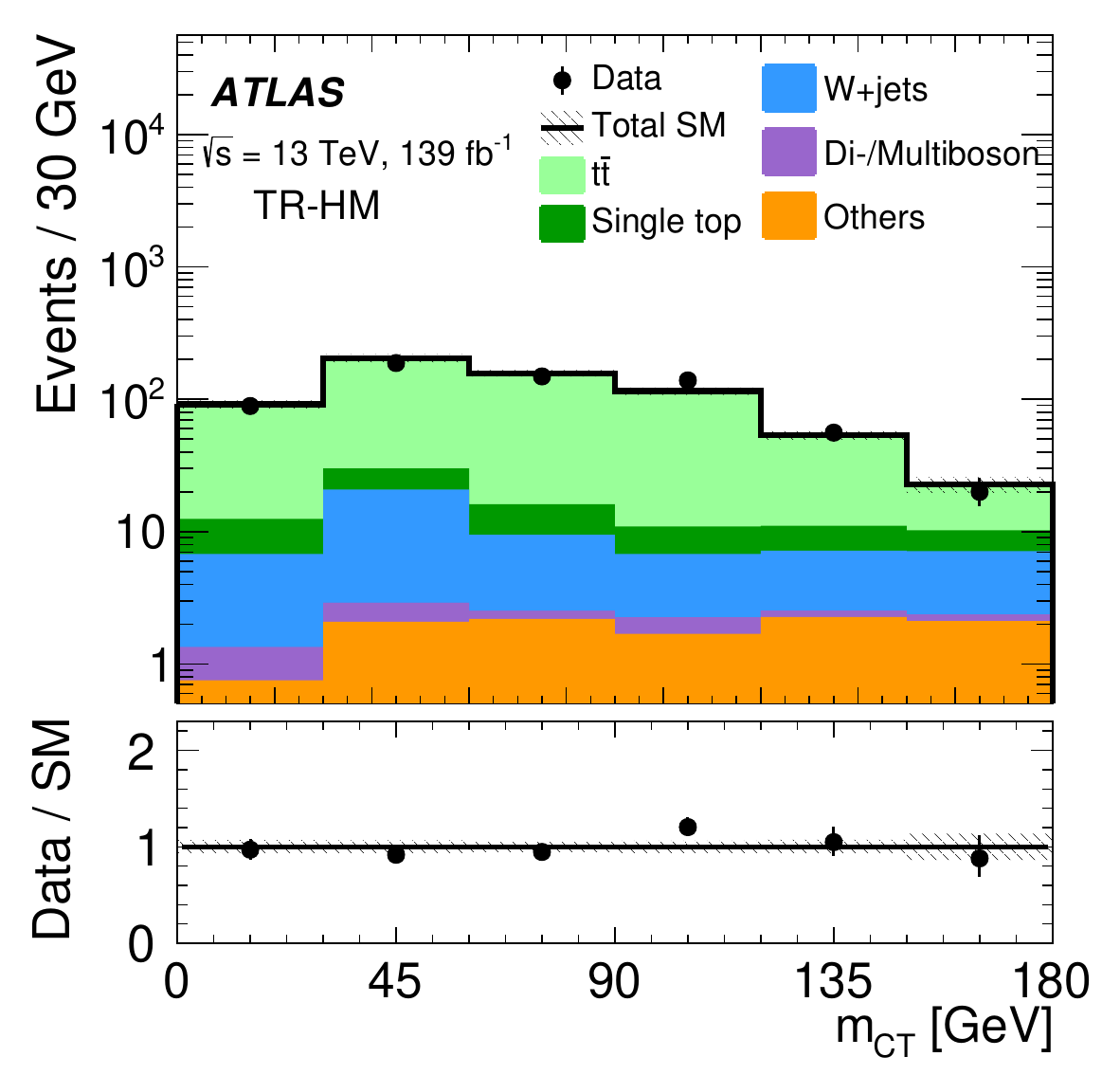}
\includegraphics[width=0.45\textwidth]{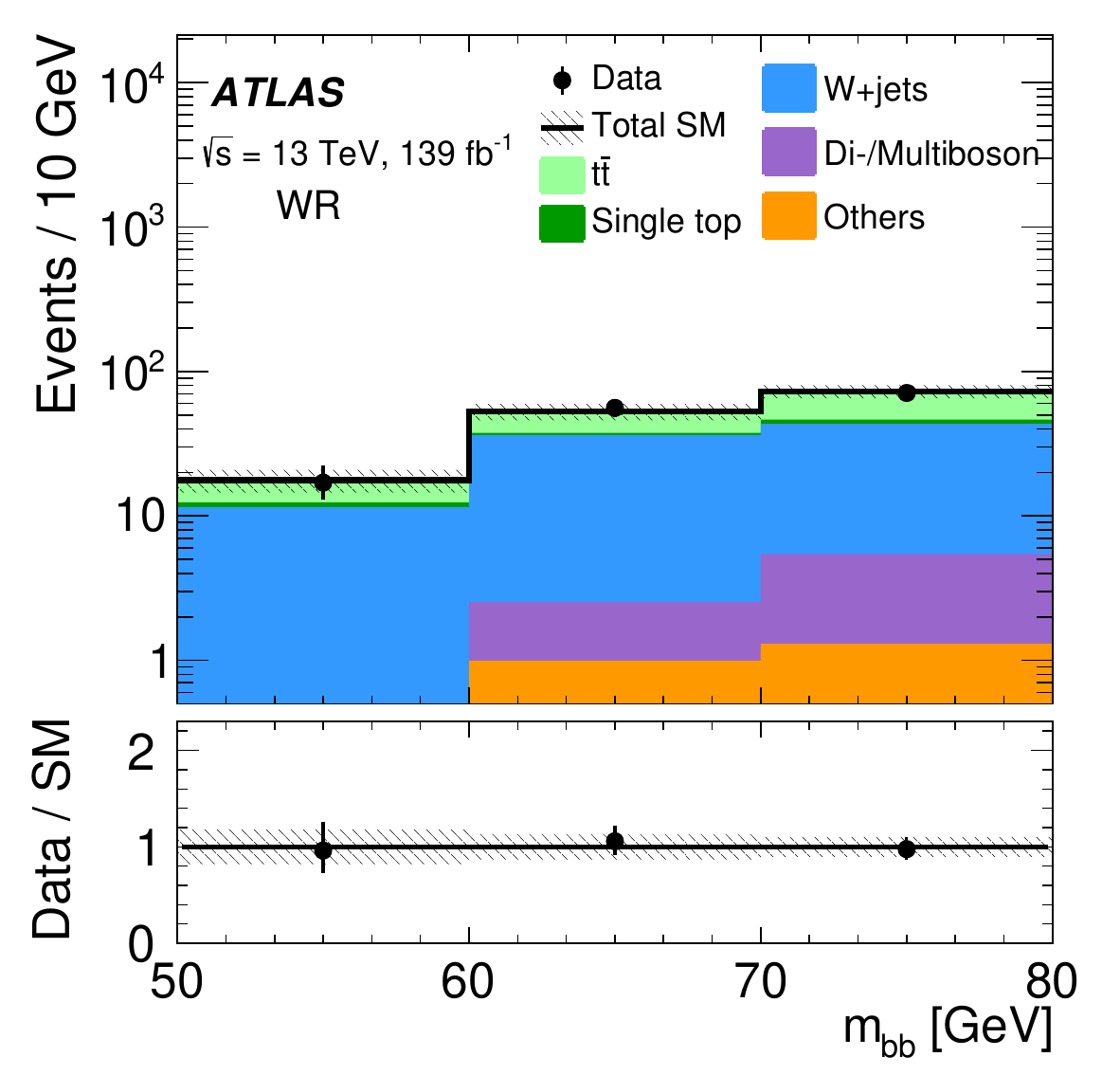}
\includegraphics[width=0.45\textwidth]{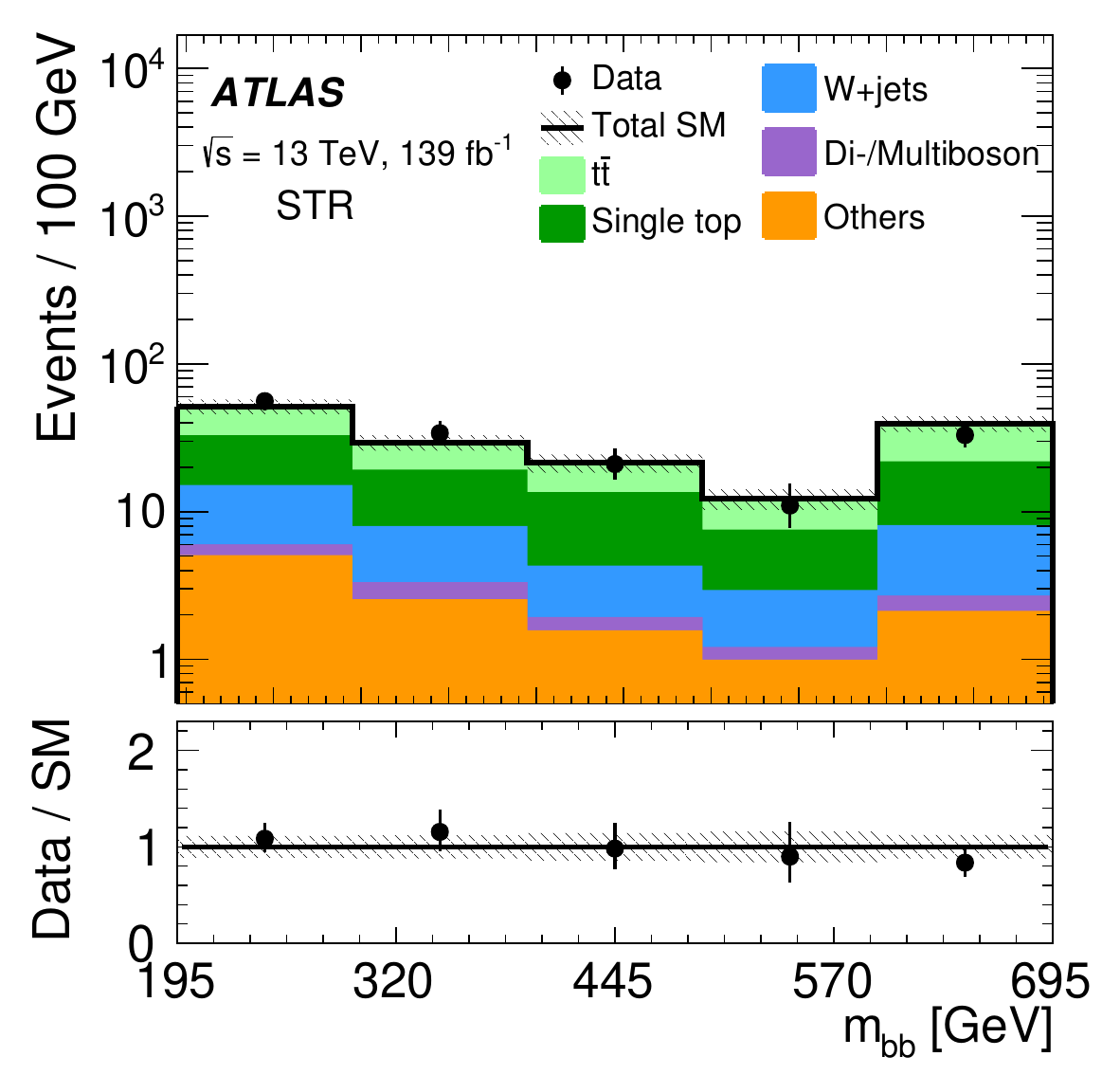}
\end{center}
\caption{The post-fit \mct\ distributions in TR-LM, TR-MM, and TR-HM are shown as well as the post-fit \mbb\ distributions in WR and STR. The uncertainty bands plotted include all statistical and systematic uncertainties. The overflow events, where present, are included in the last bin.}
\label{fig:CRs}
\end{figure}

\begin{figure}
\begin{center}
\includegraphics[width=0.45\textwidth]{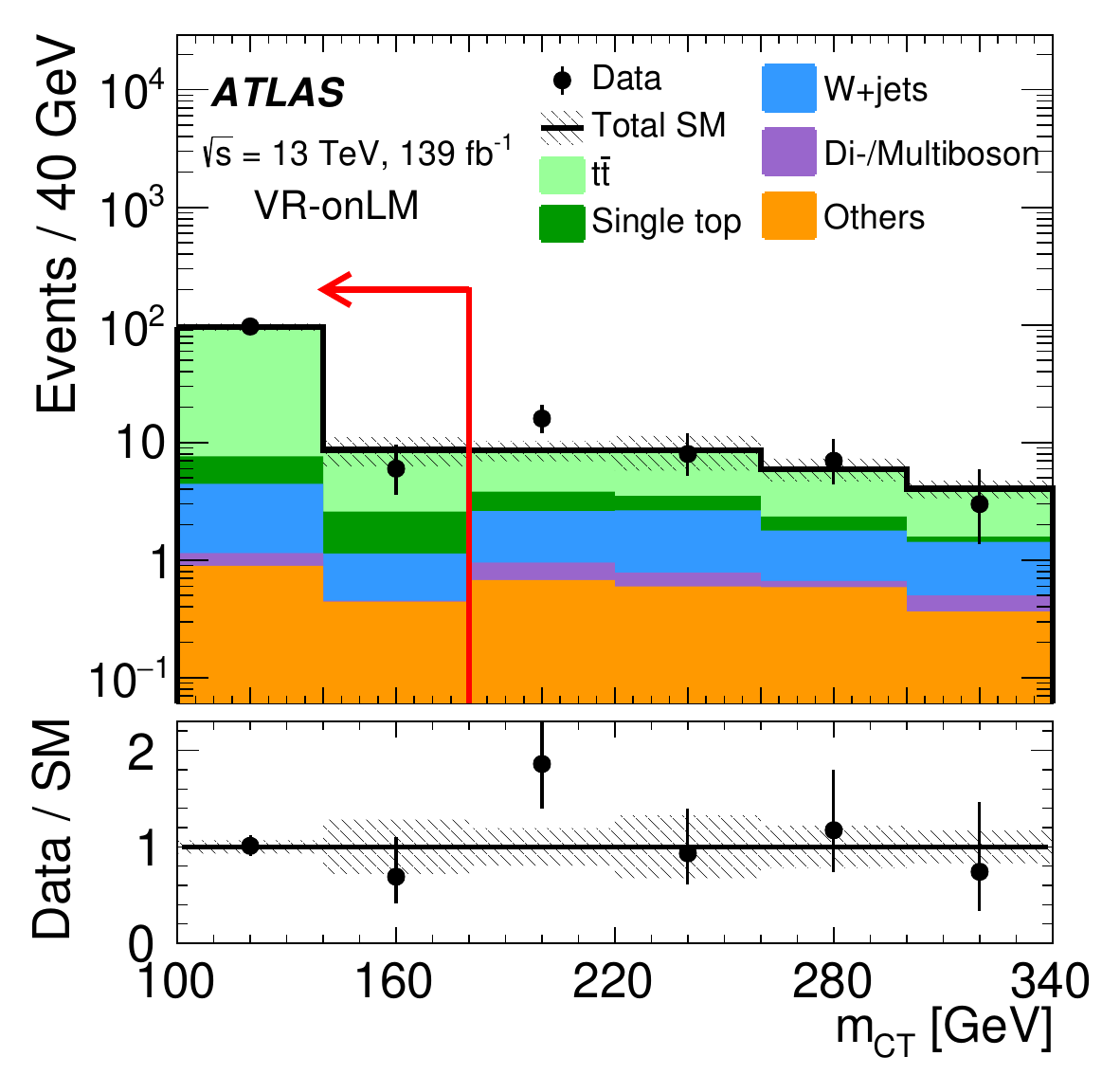}
\includegraphics[width=0.45\textwidth]{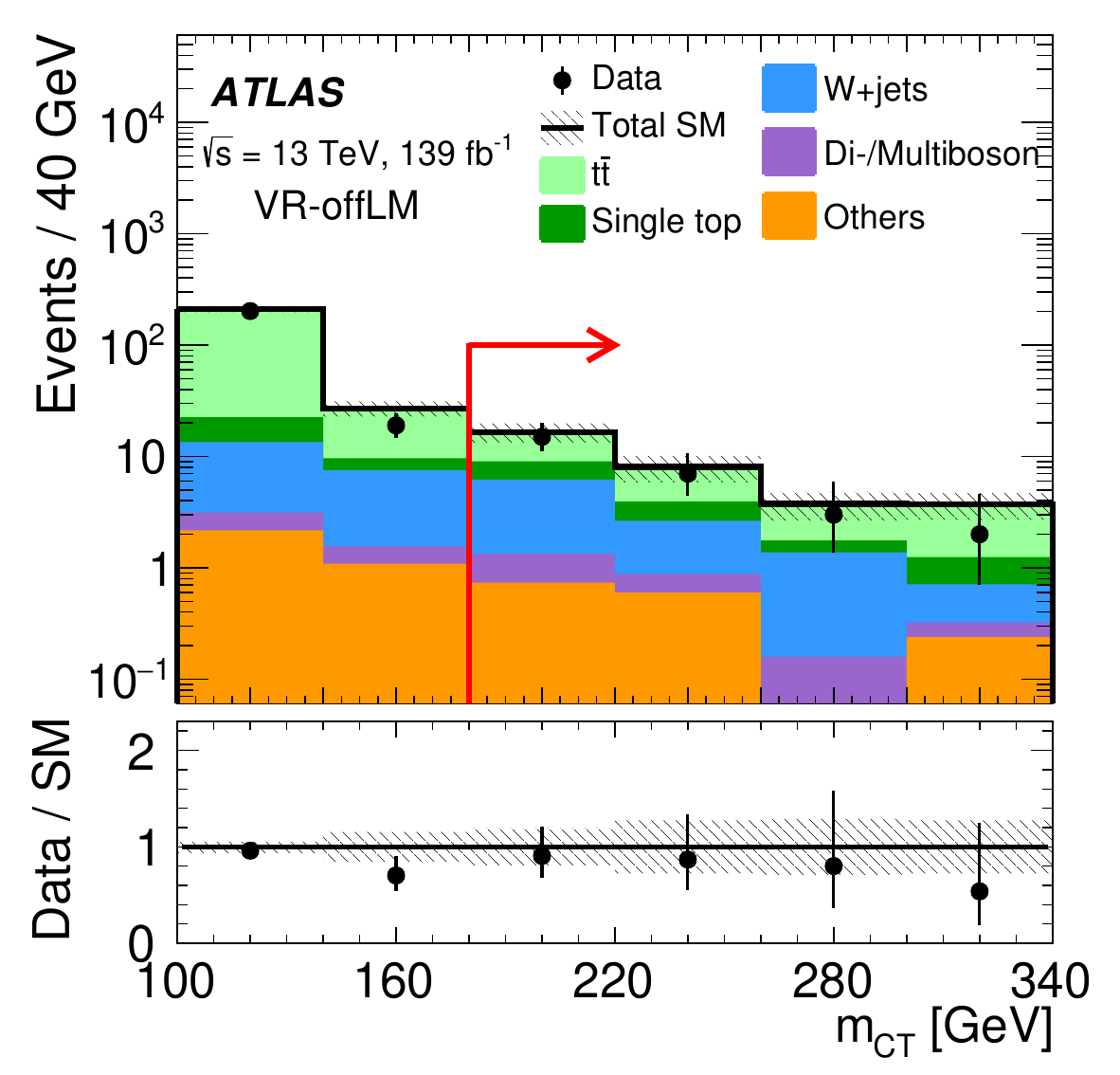}
\includegraphics[width=0.45\textwidth]{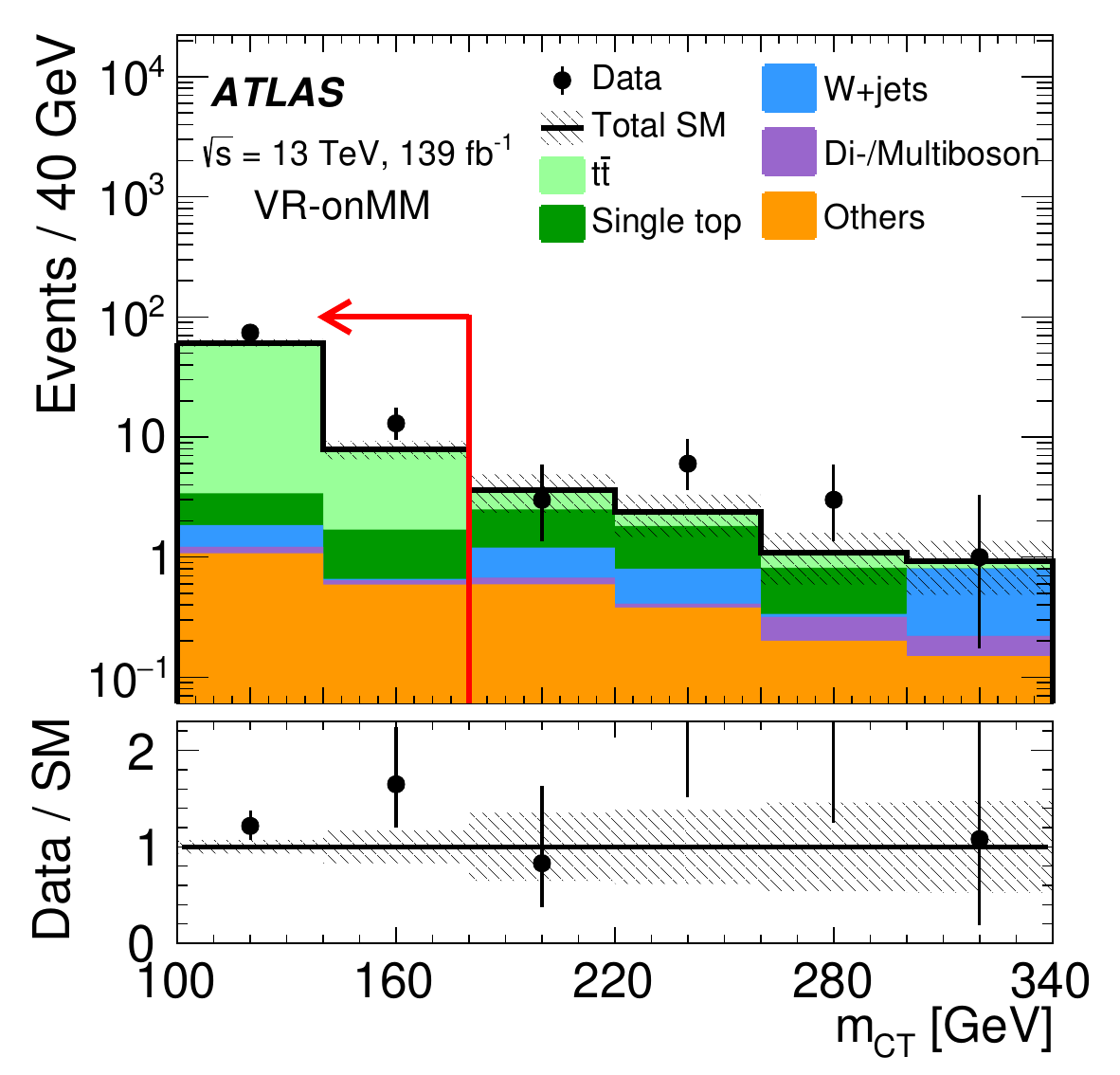}
\includegraphics[width=0.45\textwidth]{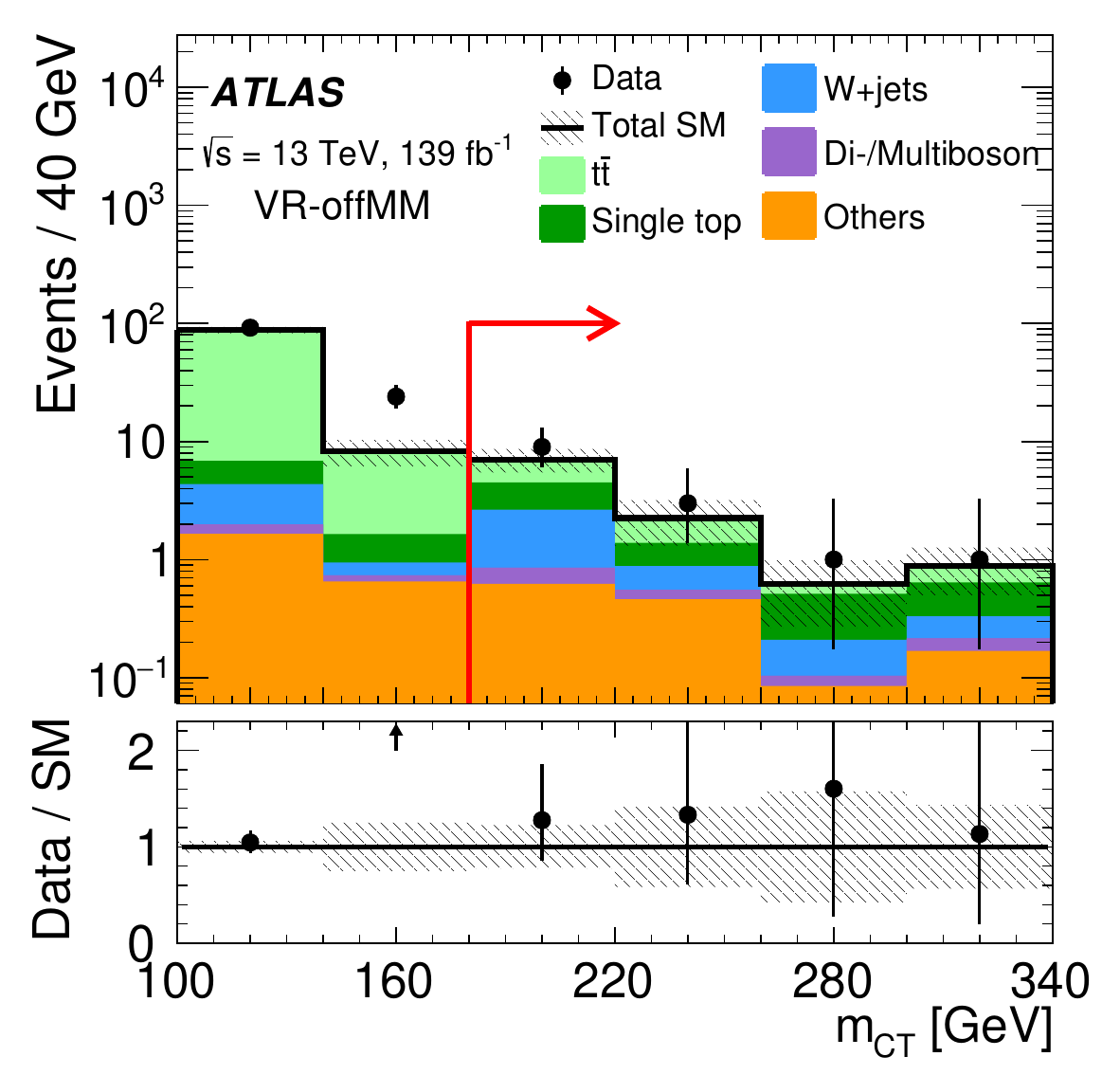}
\includegraphics[width=0.45\textwidth]{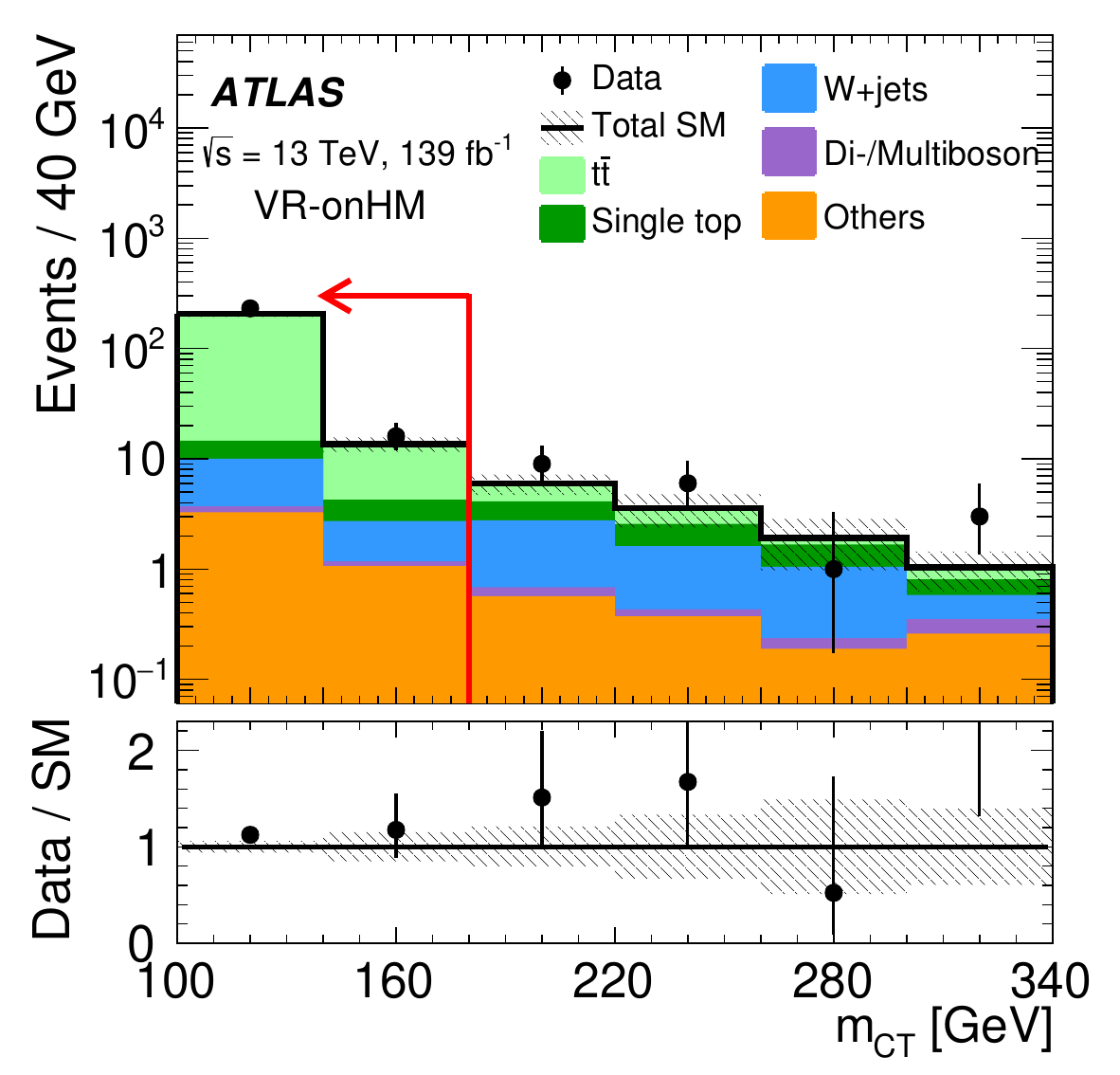}
\includegraphics[width=0.45\textwidth]{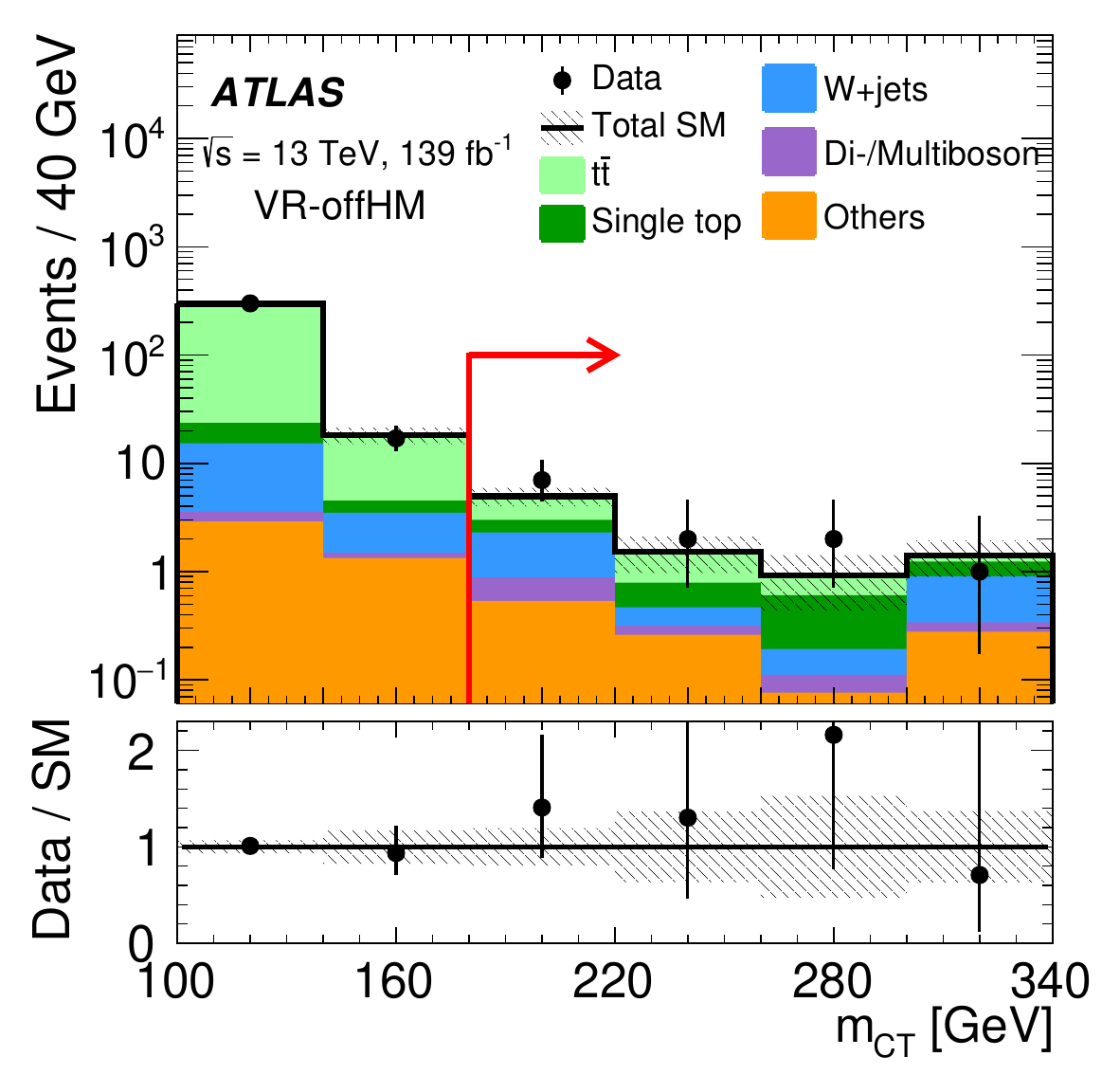}
\end{center}
\caption{The post-fit \mct\ distributions are shown in each of the validation regions  (VR-onLM, VR-onMM, VR-onHM, VR-offLM, VR-offMM, and VR-offHM) after all the selection requirements are applied other than the \mct\ cut. The uncertainty bands plotted include all statistical and systematic uncertainties. The overflow (underflow) events, where present, are included in the last (first) bin. The line with an arrow indicates the \mct\ cut used in VR selection.}
\label{fig:VRs}
\end{figure}

In Figure~\ref{fig:CRs} the post-fit \mct\ distributions in the \ttbar control regions TR-LM, TR-MM, and TR-HM are compared with the data. For the $W$ boson and single-top control regions the \mbb\ distribution is shown.
Figure~\ref{fig:VRs} shows the post-fit \mct\ distributions after all of the validation region selection requirements are applied except the \mct\ cut. 
The data and the background expectation in all validation regions agree well within two standard deviations. Therefore no further systematic uncertainty is applied on the background estimation in the signal regions.
 
\begin{figure}
\centering\includegraphics[width=\textwidth]{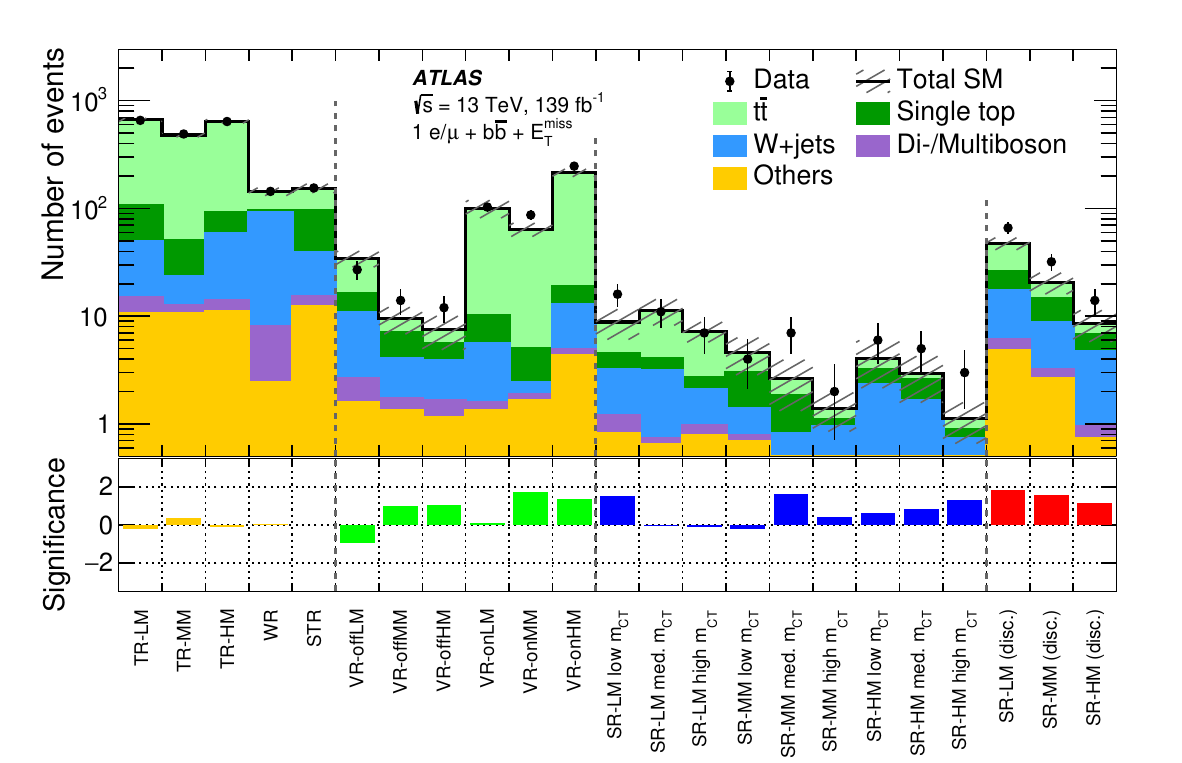}
\caption{\label{sec:results:pullplot}Comparison of the observed and expected event yields in control, validation, exclusion, and discovery signal regions. Uncertainties in the background estimates include both the statistical (in the simulated event yields) and systematic uncertainties. The bottom panel shows the significance~\cite{2008NIMPA.595..480C} of the differences between the observed and expected yields.  Not all regions shown here are statistically independent.}
\end{figure}

\begin{figure}
\includegraphics[width=0.48\textwidth]{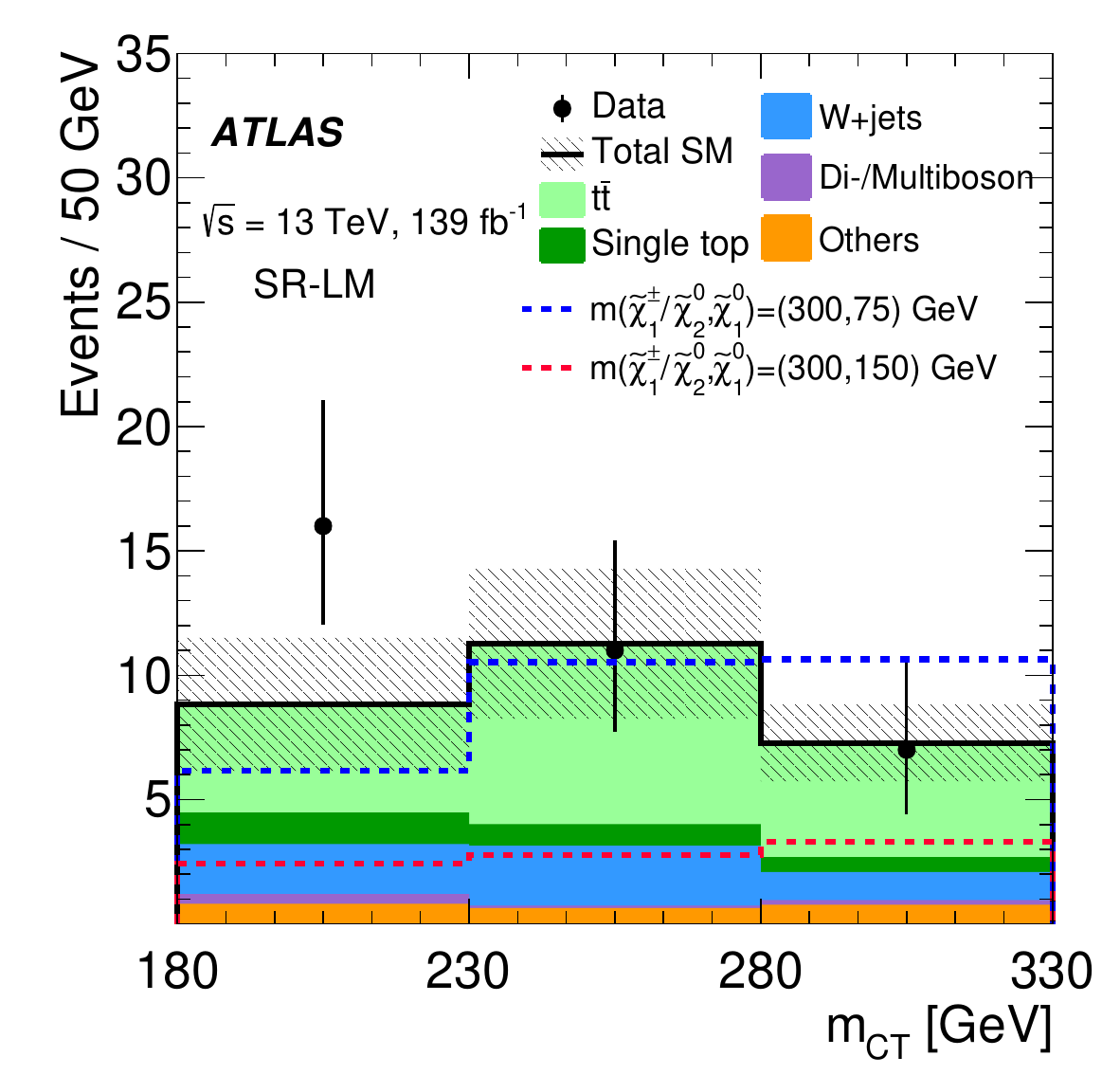}
\includegraphics[width=0.48\textwidth]{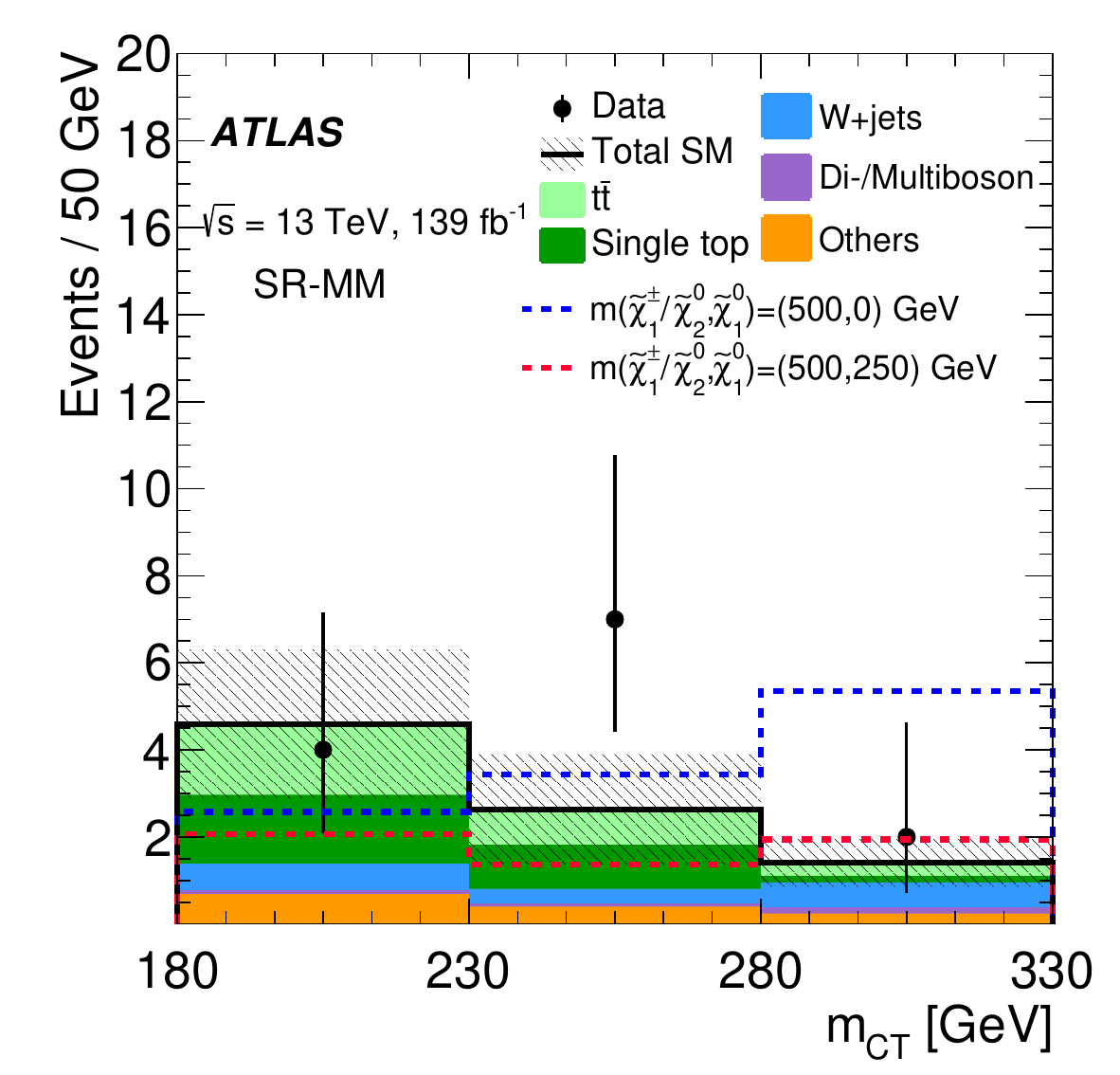}
\centering\includegraphics[width=0.48\textwidth]{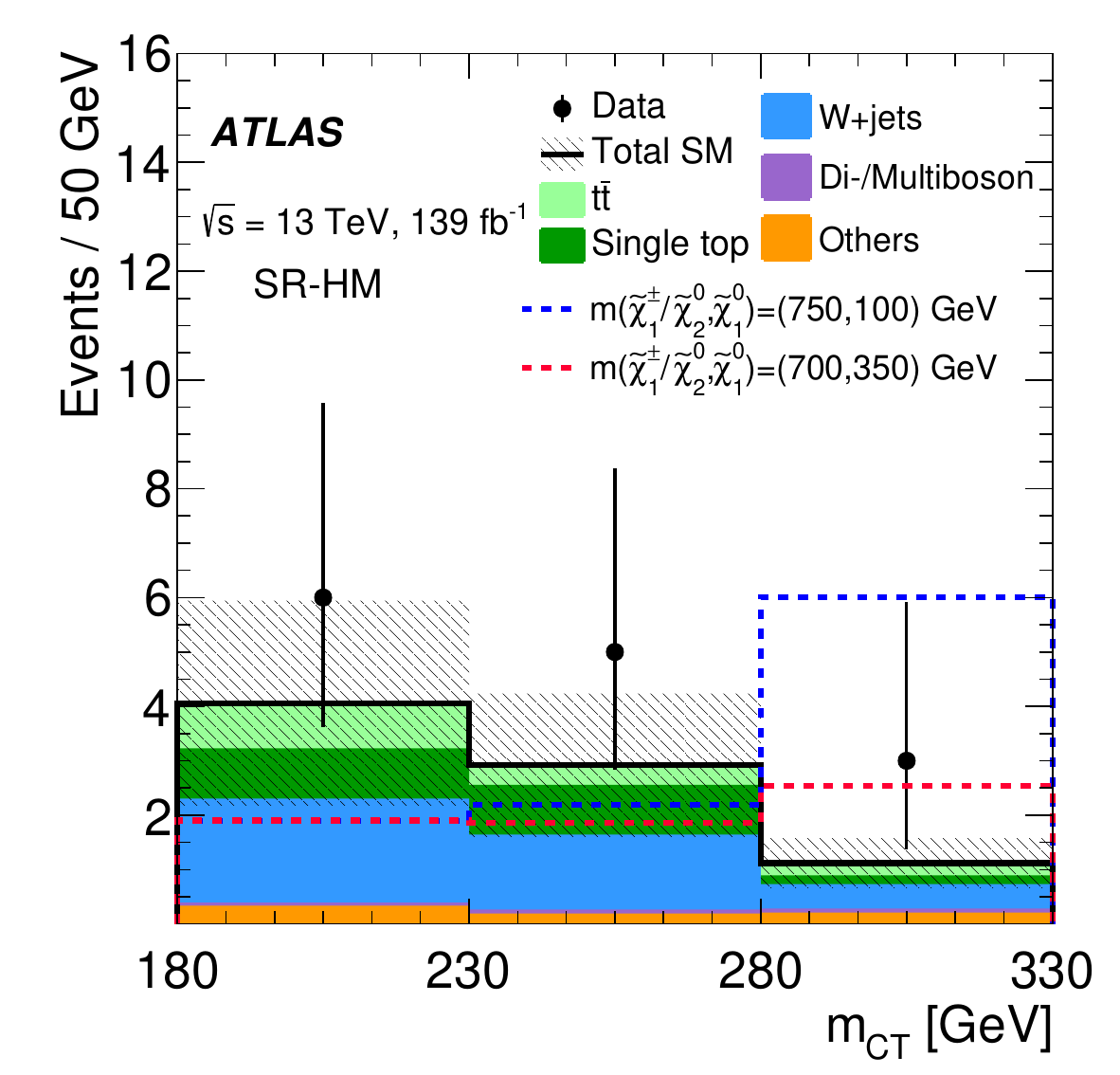}
\caption{The post-fit \mct\ distributions in the exclusion signal regions (SR-LM, SR-MM, and SR-HM). The uncertainty bands plotted include all statistical and systematic uncertainties. The dashed lines represent the benchmark signal samples. The overflow events, where present,  are included in the last bin.}
\label{fig:SRs}
\end{figure}

The compatibility of the observed and expected event yields in control, validation, exclusion, and discovery signal regions is illustrated in Figure~\ref{sec:results:pullplot}. No significant excess over the SM prediction is observed in data.
Figure~\ref{fig:SRs} shows the post-fit \mct\ distributions in SR-LM, SR-MM, and SR-HM. The uncertainty bands include all statistical and systematic uncertainties. The dashed lines represent the benchmark signal points.
 
\begin{figure}
\begin{center}
\includegraphics[width=\textwidth]{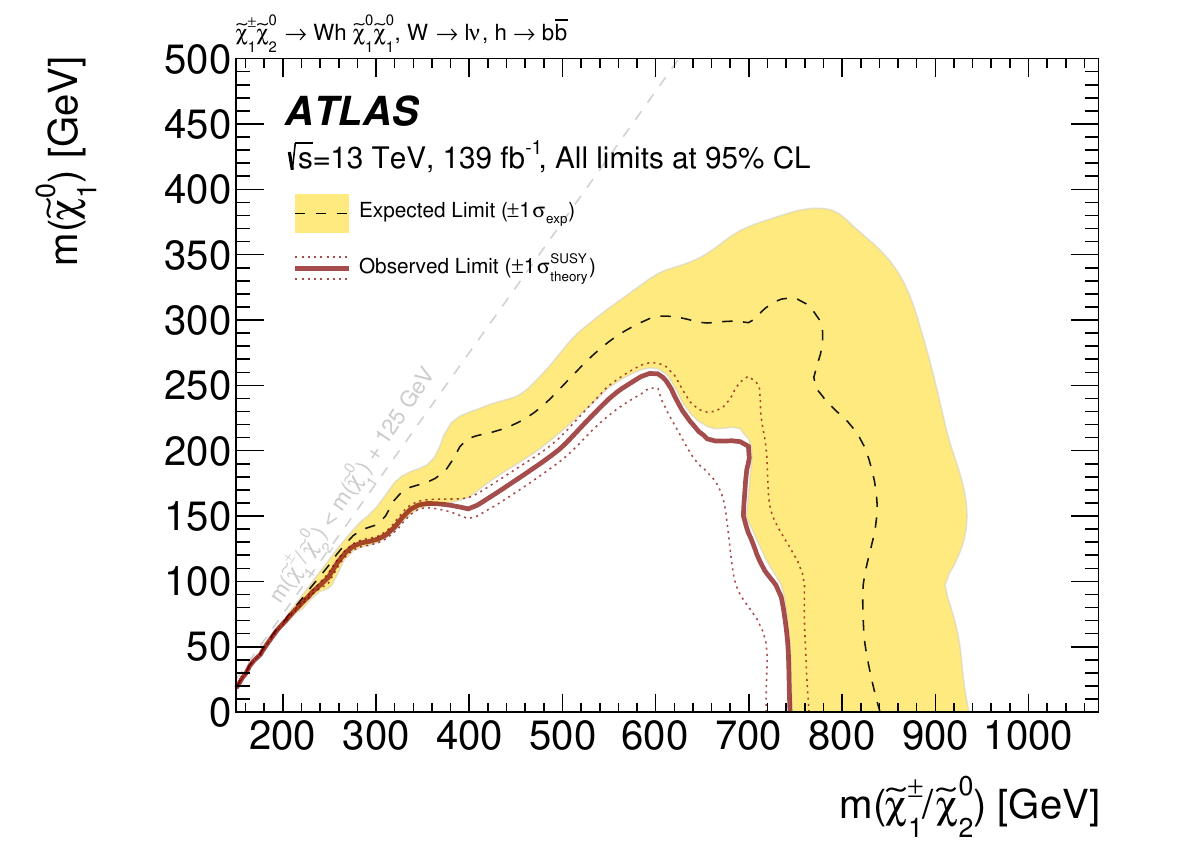}
\end{center}
\caption{\label{sec:results:limit_Wh1Lbb}Model-dependent exclusion contour at 95\% CL on the production of a chargino and a next-to-lightest neutralino. The observed limit is given by the solid line with the signal cross-section uncertainties shown by the dotted lines as indicated in the text. Expected limits are given by the dashed line with uncertainties shown by the shaded band.}
\end{figure}

Model-dependent exclusion limits at 95\% confidence level (CL) are placed on the signal model. These limits are shown as a function of the masses of the supersymmetric particles in Figure~\ref{sec:results:limit_Wh1Lbb}. They are determined interpolating between the simulated mass points, but no smoothing procedure is applied so that local fluctuations in the limit can be present. A likelihood similar to the one used in the background-only fit, but with additional terms for the SRs, is used for the calculation. The exclusion SRs thus participate in the fit and are used to constrain normalisation and nuisance parameters. A signal is allowed in this likelihood in both the CRs and SRs. The VRs are not used in the fit. The
$\mathrm{CL_s}$ method~\cite{Read:2002hq} is used to derive the confidence level of the exclusion for a particular signal model;
signal models with a $\mathrm{CL_s}$ value below 0.05 are excluded at 95\% CL. The uncertainties in the observed limit are calculated by varying the cross-section for the signal up and down by its uncertainty. Due to a modest excess observed in some bins of the exclusion signal regions, the observed limit is weaker than the expected limit and extends up to about $740$~\GeV\ in $m(\chinoonepm/\ninotwo)$ for massless \ninoone\ and up to $m(\chinoonepm/\ninotwo) = 600$~\GeV\ for $m(\ninoone) = 250$~\GeV. Benefiting from the increased integrated luminosity and the improved two-dimensional fit procedure, the current observed limit exceeds the previous ATLAS limit by about $200$~\GeV\ in $m(\chinoonepm/\ninotwo)$ for a massless \ninoone.
 
Table~\ref{upperlimit_toys} summarises the observed ($S_{\mathrm{obs}}$) and expected ($S_{\mathrm{exp}}$) 95\% CL upper limits on the number of signal events and on the observed visible cross-section,
$\langle\epsilon{\mathrm{ \sigma}}\rangle_{\mathrm{ obs}}^{95}$,
for each of the three cumulative discovery signal regions. These cumulative signal regions are those defined to test for the presence of any beyond-the-Standard-Model (BSM) physics processes, where in every case the upper bound on \mt\ is also removed. Upper limits on contributions from new physics processes are estimated using the so-called `model-independent fit', where a generic BSM process is assumed to contribute only to the SR and not to the CRs, thus giving a conservative background estimate in the SR.
When normalised to the integrated luminosity of the data sample, the results can be interpreted as corresponding to observed upper limits
$\langle\epsilon{\mathrm{ \sigma}}\rangle_{\mathrm{ obs}}^{95}$,
defined as the product of the production cross-section, the acceptance, and the selection efficiency of a
BSM signal. The $p_{0}$ values, which represent the probability of the SM background alone to fluctuate to the
observed number of events or higher, are also provided. All numbers are calculated from pseudo-experiments.

\begin{table}
\centering
\setlength{\tabcolsep}{0.0pc}
\begin{tabular*}{\textwidth}{@{\extracolsep{\fill}}lcccccc}
\noalign{\smallskip}\hline\noalign{\smallskip}
{\textbf{Signal Region}}                        & $\langle\epsilon{\mathrm{ \sigma}}\rangle_{\mathrm{ obs}}^{95}$[fb]  &  $S_{\mathrm{ obs}}^{95}$  & $S_{\mathrm{ exp}}^{95}$ & $\textrm{CL}_{\textrm{B}}$ & $p_{0}$ & $Z$  \\
\noalign{\smallskip}\hline\noalign{\smallskip}
SR-LM$\mathrm{(disc.)}$    & $0.26$ &  $36.8$ & $ { 20.0 }^{ +8.0 }_{ -5.4 }$ & $0.97$ & $ 0.03$&$1.88$ \\%
SR-MM$\mathrm{(disc.)}$    & $0.18$ &  $24.8$ & $ { 15.3 }^{ +6.2 }_{ -4.6 }$ & $0.94$ & $ 0.06$&$1.54$ \\%
SR-HM$\mathrm{(disc.)}$    & $0.11$ &  $14.7$ & $ { ~~9.7 }^{ +3.3 }_{ -2.7 }$ & $0.89$ & $ 0.10$&$1.30$ \\%
 
\noalign{\smallskip}\hline\noalign{\smallskip}
\end{tabular*}
\caption[Breakdown of upper limits.]{
Left to right: 95\% CL upper limits on the visible cross-section
($\langle\epsilon\sigma\rangle_{\mathrm{ obs}}^{95}$) and on the number of
signal events ($S_{\mathrm{ obs}}^{95}$ ).
The third column ($S_{\mathrm{ exp}}^{95}$) shows the expected 95\% CL upper limit (and its $\pm 1\sigma$ excursions)
on the number of signal events if no BSM signal is present.
The last three columns
indicate the $\textrm{CL}_{\textrm{B}}$ value, i.e. the confidence level observed for
the background-only hypothesis, the discovery $p$-value ($p_{0}$) and the significance $Z$~\cite{2008NIMPA.595..480C}.
\label{upperlimit_toys}}
\end{table}
% End of text imported from the .//section/result.tex input file

\FloatBarrier

\section{Conclusion}
 
The results of a search for electroweakino pair production $pp \rightarrow \tilde\chi^\pm_1 \tilde\chi^0_2$ in which the chargino ($\tilde\chi^\pm_1$) decays into a $W$ boson and the lightest neutralino ($\tilde\chi^0_1$), while the heavier neutralino ($\tilde\chi^0_2$) decays into the Standard Model 125~\GeV\ Higgs boson and a second $\tilde\chi^0_1$ are presented.
The analysis is performed using \(pp\) collisions provided by the LHC at a centre-of-mass energy of 13~\TeV.
Data collected with the ATLAS detector between 2015 and 2018 are used, corresponding to an integrated luminosity of $139~\mathrm{fb}^{-1}$.
No significant deviation from the expected Standard Model background is observed.
Limits are set on the direct production of the electroweakino in simplified models. Masses of $\tilde{\chi}^{\pm}_{1}/\tilde{\chi}^{0}_{2}$ up to 740~\GeV\ are excluded at 95\% confidence level for a massless $\tilde{\chi}^{0}_{1}$. The current search improves on the previous ATLAS limit by about $200$~\GeV\ in $m(\chinoonepm/\ninotwo)$ for a massless \ninoone.

\section*{Acknowledgements}
 
% The next lines are included from the .//acknowledgements/Acknowledgements.tex input file

We thank CERN for the very successful operation of the LHC, as well as the
support staff from our institutions without whom ATLAS could not be
operated efficiently.
 
We acknowledge the support of ANPCyT, Argentina; YerPhI, Armenia; ARC, Australia; BMWFW and FWF, Austria; ANAS, Azerbaijan; SSTC, Belarus; CNPq and FAPESP, Brazil; NSERC, NRC and CFI, Canada; CERN; CONICYT, Chile; CAS, MOST and NSFC, China; COLCIENCIAS, Colombia; MSMT CR, MPO CR and VSC CR, Czech Republic; DNRF and DNSRC, Denmark; IN2P3-CNRS and CEA-DRF/IRFU, France; SRNSFG, Georgia; BMBF, HGF and MPG, Germany; GSRT, Greece; RGC and Hong Kong SAR, China; ISF and Benoziyo Center, Israel; INFN, Italy; MEXT and JSPS, Japan; CNRST, Morocco; NWO, Netherlands; RCN, Norway; MNiSW and NCN, Poland; FCT, Portugal; MNE/IFA, Romania; MES of Russia and NRC KI, Russia Federation; JINR; MESTD, Serbia; MSSR, Slovakia; ARRS and MIZ\v{S}, Slovenia; DST/NRF, South Africa; MINECO, Spain; SRC and Wallenberg Foundation, Sweden; SERI, SNSF and Cantons of Bern and Geneva, Switzerland; MOST, Taiwan; TAEK, Turkey; STFC, United Kingdom; DOE and NSF, United States of America. In addition, individual groups and members have received support from BCKDF, CANARIE, Compute Canada and CRC, Canada; ERC, ERDF, Horizon 2020, Marie Sk{\l}odowska-Curie Actions and COST, European Union; Investissements d'Avenir Labex, Investissements d'Avenir Idex and ANR, France; DFG and AvH Foundation, Germany; Herakleitos, Thales and Aristeia programmes co-financed by EU-ESF and the Greek NSRF, Greece; BSF-NSF and GIF, Israel; CERCA Programme Generalitat de Catalunya and PROMETEO Programme Generalitat Valenciana, Spain; G\"{o}ran Gustafssons Stiftelse, Sweden; The Royal Society and Leverhulme Trust, United Kingdom.
 
The crucial computing support from all WLCG partners is acknowledged gratefully, in particular from CERN, the ATLAS Tier-1 facilities at TRIUMF (Canada), NDGF (Denmark, Norway, Sweden), CC-IN2P3 (France), KIT/GridKA (Germany), INFN-CNAF (Italy), NL-T1 (Netherlands), PIC (Spain), ASGC (Taiwan), RAL (UK) and BNL (USA), the Tier-2 facilities worldwide and large non-WLCG resource providers. Major contributors of computing resources are listed in Ref.~\cite{ATL-GEN-PUB-2016-002}.
 
% End of text imported from the .//acknowledgements/Acknowledgements.tex input file

\clearpage

\printbibliography

\clearpage % ATLAS Collaboration author list
% Reference date of SUSY-2019-08 is 2019-06-28
% Author list last updated on date 23-APR-20
% Data extracted on 23-Apr-2020 for paper reference SUSY-2019-08
% at 12:01pm
 
\begin{flushleft}
\hypersetup{urlcolor=black}
{\Large The ATLAS Collaboration}

\bigskip

\AtlasOrcid[0000-0002-6665-4934]{G.~Aad}$^\textrm{\scriptsize 102}$,    
\AtlasOrcid[0000-0002-5888-2734]{B.~Abbott}$^\textrm{\scriptsize 129}$,    
\AtlasOrcid{D.C.~Abbott}$^\textrm{\scriptsize 103}$,    
\AtlasOrcid[0000-0002-2788-3822]{A.~Abed~Abud}$^\textrm{\scriptsize 71a,71b}$,    
\AtlasOrcid{K.~Abeling}$^\textrm{\scriptsize 53}$,    
\AtlasOrcid{D.K.~Abhayasinghe}$^\textrm{\scriptsize 94}$,    
\AtlasOrcid[0000-0002-8496-9294]{S.H.~Abidi}$^\textrm{\scriptsize 167}$,    
\AtlasOrcid[0000-0002-8279-9324]{O.S.~AbouZeid}$^\textrm{\scriptsize 40}$,    
\AtlasOrcid{N.L.~Abraham}$^\textrm{\scriptsize 156}$,    
\AtlasOrcid[0000-0001-5329-6640]{H.~Abramowicz}$^\textrm{\scriptsize 161}$
% ...
\bigskip
\\

% ...
$^{40}$Niels Bohr Institute, University of Copenhagen, Copenhagen; Denmark.\\
% ...
$^{53}$II. Physikalisches Institut, Georg-August-Universit\"{a}t G\"ottingen, G\"ottingen; Germany.\\
% ...
$^{71}$$^{(a)}$INFN Sezione di Pavia;$^{(b)}$Dipartimento di Fisica, Universit\`a di Pavia, Pavia; Italy.\\
% ...
$^{94}$Department of Physics, Royal Holloway University of London, Egham; United Kingdom.\\
% ...
$^{102}$CPPM, Aix-Marseille Universit\'e, CNRS/IN2P3, Marseille; France.\\
$^{103}$Department of Physics, University of Massachusetts, Amherst MA; United States of America.\\
% ...
$^{129}$Homer L. Dodge Department of Physics and Astronomy, University of Oklahoma, Norman OK; United States of America.\\
% ...
$^{156}$Department of Physics and Astronomy, University of Sussex, Brighton; United Kingdom.\\
% ...
$^{161}$Raymond and Beverly Sackler School of Physics and Astronomy, Tel Aviv University, Tel Aviv; Israel.\\
% ...
$^{167}$Department of Physics, University of Toronto, Toronto ON; Canada.\\
% ...
$^{a}$ Also at Borough of Manhattan Community College, City University of New York, New York NY; United States of America.\\
$^{b}$ Also at CERN, Geneva; Switzerland.\\
% ...

\end{flushleft}

% Created with Glance <Atlas.Glance@cern.ch>

\end{document}